\documentclass[aps,prl,reprint,superscriptaddress,nofootinbib,showpacs,preprintnumbers]{cernrep}
\pdfoutput=1

\usepackage{graphicx} 
\usepackage{atlasphysics}
\usepackage{subfigure}
\usepackage{longtable}
\usepackage{multirow}
\usepackage{textcomp}
\usepackage{units}
\usepackage{lineno}
\usepackage{rotating}
\usepackage{amssymb}
\usepackage{amsmath}
\usepackage{array} 
\usepackage[colorlinks=true,linkcolor=blue,citecolor=blue]{hyperref}
\usepackage[font=large]{caption}
\DeclareGraphicsExtensions{.pdf}
\begin{document}

\begin{titlepage}

\begin{flushleft}
  DESY 18-021
\end{flushleft}

\noindent
\vspace*{2.5cm}

  \begin{center}
    \begin{Large}
      

      {\bfseries QCD analysis of the ATLAS and CMS \pmb {$W^{\pm}$} and \pmb{$Z$}
    cross-section measurements and implications for the strange sea density}

   \vspace*{2cm}

A.M.~Cooper-Sarkar$^a$,
K.~Wichmann$^b$

\end{Large}
\end{center}


{\protect \hskip 0.cm} $^a$ Physics Department, University of Oxford, Oxford OX1 3RH, United Kingdom
 
\vspace*{-.12cm}
{\protect \hskip 0.cm} $^b$ Deutsches Elektronen Synchrotron DESY, Hamburg 22607, Germany

\vspace*{1cm}

\begin{abstract} \noindent

  In the present paper, the ATLAS inclusive $W^{\pm}$ and $Z$ boson production data are analysed
  together with the CMS inclusive $W^{\pm}$ and $Z$ boson 
production data to investigate any possible tensions between the data sets and 
to determine the strange sea fraction, within the framework of a parton distribution 
function fit at next-to-next-to leading order in perturbative QCD.

\end{abstract}

\vspace*{1.5cm}

\end{titlepage}

\newpage
~~
\newpage

\renewcommand{\arraystretch}{1.35}
\section{Introduction}
Measurements that probe the parton distribution 
functions (PDFs) of the proton have traditionally 
been made in deep inelastic scattering (DIS)
experiments. A very broad range of resolving power, as characterised by $Q^2$,
 the negative four-momentum transfer squared, and of Bjorken $x$, 
(which is interpreted as the fraction of the proton's momentum taken by the 
struck parton) has been measured. The state of the art interpretations of 
these data use
next-to-next-to leading order (NNLO) calculations in perturbative Quantum 
Chromodynamics (QCD) to determine the parton distribution functions.

The region of low Bjorken $x$, $x < \sim 0.01$, is primarily 
constrained by the precise measurement of 
the proton structure function $F_2(x,Q^2)$ at 
HERA~\cite{1506.06042}, which  determines a specific
combination of light quark and anti-quark distributions.
However, the flavour composition of the total light sea, 
is not well determined using the HERA data alone.
In particular, little is known about the strange sea.
Historically, the strange sea was related to the
light quark sea by an $x$-independent 
fraction~\cite{JimenezDelgado:2008hf, HERA:2009wt, Adloff:2000qk},
such as $\bar{s}(x)= r_s \bar{d}(x)$, and it is often assumed that the strange 
sea is suppressed such that 
$r_s\sim 0.5$. The evidence for this suppression 
comes from dimuon production in charged current data from the
NuTeV~\cite{PhysRevD.74.012008}, 
CCFR~\cite{Goncharov:2001qe} and NOMAD~\cite{Samoylov:2013} neutrino scattering experiments.
These data provide  constraints at larger $x$ 
on the strange and also the anti-strange densities through the subprocesses
$W^+ s \rightarrow c$ and $W^- \overline{s} \rightarrow \overline{c}$. 
However, the interpretation of these data is sensitive to uncertainties 
from charm fragmentation and nuclear corrections and when these data have 
been used 
to determine parameters of the strange sea quark distribution no consensus, between PDF fitting groups,
on the level of suppression as a function of $x$ has emerged. 

The high precision measurements of the inclusive 
$W$ and $Z$ boson cross section
at the LHC recently performed by the ATLAS collaboration provide new 
constraints on the strange quark density in the low-$x$ regime~\cite{1612.03016} 
and a corresponding set of NNLO PDFs, the ATLASepWZ16 PDFs.
Specifically, the data support the hypothesis of a symmetric composition of the
light quark sea at low-$x$ for both high and low scales. This observation was 
supported by analysis of the ATLAS $W+c$ data~\cite{atlaswcharm} . 
However, the CMS $W+c$ data~\cite{cmswcharm}  
favour a somewhat smaller strangeness. Since the analysis of the $W+c$ data involve 
assumptions on charm jet fragmentation and hadronisation it is interesting to investigate if this 
disagreement is present for the inclusive Drell-Yan (DY) data of ATLAS and CMS. 
The Drell-Yan process and DIS are theoretically the best understood processes. 
In the present paper, the ATLAS inclusive $W$ and $Z$ differential cross section 
data are analysed together with the CMS 
inclusive $W$ and $Z$ differential cross section data 
to investigate any possible tensions between these data sets and 
to determine the strange sea fraction. This is done by performing 
 a parton distribution function analysis in NNLO QCD using the  
 inclusive deep inelastic scattering data from HERA 
 jointly with the ATLAS and CMS inclusive Drell-Yan data.

 The results are presented in terms of the ratio $(s +\bar{s})/(\bar{u} +\bar{d})$ as a function of $x$
 for both low and high-$Q^2$ scales. This ratio is unity if strange quarks are not 
suppressed in relation to light quarks and is $\sim 0.5$ for the conventional level of 
suppression. However, two points should be noted, firstly the experimental evidence for suppression 
is centred at higher $x$, $x>\sim 0.1$ and modest $Q^2\sim 20$\,GeV$^2$, whereas the 
sensitivity of the LHC data is at $x\sim 0.01$. Secondly, whatever the level of 
suppression at low $Q^2$, the fraction of strangeness will grow as $Q^2$ increases due to flavour 
blind $g \to q \bar{q}$ splitting, such that the strangeness ratio tends asymptotically to unity for 
large $Q^2$. Thus at LHC scales the difference between suppression and non-suppression 
at the starting scale requires precision measurements.

\section{Input Data Sets}

The final combined $e^{\pm}p$
cross section measurements of HERA~\cite{1506.06042} 
cover the kinematic range of  $Q^2$
from $0.045$\,GeV$^2$ to $50000$\,GeV$^2$ and
of Bjorken $x$ from $0.65$ down to $6\times 10^{-7}$. There are 169 correlated sources of uncertainty
and total uncertainties are below 1.5$\%$ over the $Q^2$ range $3 < Q^2 <500$\,GeV$^2$
and below 3$\%$ up to $Q^2=3000$\,GeV$^2$.

The ATLAS $W^{\pm}$ differential 
cross sections are based on combined data in both electron and muon decay channels
as functions of the $W$ decay lepton ($e$,\,$\mu$)
pseudo-rapidity, $\eta$, with an experimental precision of $0.6-1.0\%$.
The $Z/\gamma^*$ boson rapidity, $y$, has been measured in three $Z$ mass ranges:
$46<M_Z<66$\,GeV; $66<M_Z<116$\,GeV 
and $116<M_Z<136$\,GeV  and in central and forward rapidity ranges, with an 
experimental precision of $0.4\%$ for central rapidity and $2.3\%$ for forward rapidity. 
The absolute normalisation of the $W,Z$ cross sections is known to within $1.8$\,\%. 
There are 131 sources of correlated systematic uncertainty. 

The CMS inclusive data come from several different analyses. The $W$ data are available 
for 7\,TeV~\cite{1312.6283} and 8\,TeV~\cite{1603.01803} collisions and are used as $W$-asymmetry data
for 7\,TeV and as separate $W^+$ and $W^-$ data for 8\,TeV. 
These data are presented as a function of muon
pseudo-rapidity with an experimental precision of $ \approx 1\%$.
The correlations are supplied as a systematic correlation matrix for the 7\,TeV data
and as full covariance matrices for the 8\,TeV data.
The $Z$ data come as double differential Drell-Yan measurements as functions of 
dimuon rapidity for different dimuon mass bins for 7\,TeV~\cite{1310.7291}
and as functions of di-lepton (combined electron and muon channels) rapidity in the same rapidity bins
for 8\,TeV~\cite{1412.1115}.
A typical experimental accuracy for these measurements is between 1\,\% and 3\,\%.
The correlations are provided as full covariance matrices. 
Only the 7\,TeV~\cite{1310.7291} $Z$ data are used for the main analysis, since the covariance 
matrix of the 8\,TeV $Z$ data results in an unreasonably large $\chi^2$ for the fits.
The same effect has been found by other PDF fitting analyses~\cite{nnpdf31}.

In principle, there can be correlations between the CMS and ATLAS data arising from: 
the use of common Monte Carlos when modelling backgrounds and signals; 
common PDF uncertainties on the measured cross 
sections and from luminosity. The former two sources of correlation are impossible to estimate 
since the identity of sources of systematic correlation is lost in the construction of a 
covariance matrix. However, these sources of uncertainty are small (per mille) 
contributions to the total uncertainty. The luminosity uncertainty is larger,
$\sim 1.8\%$ and $\sim 2.6\%$ for ATLAS and CMS 7 TeV data, respectively,
however the need to consider correlation with CMS luminosity is avoided since the CMS 7 TeV 
$Z$ data is normalised and the  $W$ data is used as a $W$-asymmetry. 
The CMS 8 TeV data come from a different running period, such that luminosity is not correlated.   

\section{Theoretical framework}

The present QCD analysis uses the 
xFitter framework~\cite{HERAFitter,HERA:2009wt,Aaron:2009kv}.
The light quark coefficient functions are calculated to NNLO 
as implemented in QCDNUM~\cite{Botje:2010ay}. 
The contributions of heavy quarks are calculated in the general-mass
variable-flavor-number scheme of Refs.~\cite{Thorne:1997ga,Thorne:2006qt}.
The renormalisation and factorisation scales for the DIS processes are taken as 
$\mu_r=\mu_f=\sqrt(Q^2)$. The program MINUIT~\cite{minuit} is used for the minimisation.
Each step is cross-checked with an independent fit program~\cite{zeusfitter}.

For all data sets the xFitter package~\cite{xFittermanual} uses the APPLGRID code~\cite{Carli:2010rw}
interfaced to the MCFM program~\cite{Campbell:2010ff}
for fast calculation of the differential $W$ and $Z/\gamma^*$
boson cross sections at NLO in QCD and LO in EW. A $K$-factor
technique is used to correct from NLO to NNLO predictions in QCD. 
For the CMS $W$ and $Z$ data the NNLO corrections are calculated with
the FEWZ3.1~\cite{ref2425of161203016} program,
using NNLO CT10~\cite{ct10} PDFs. The lowest di-lepton invariant mass
bin $ 20 \to 30$ GeV is not used in the analysis since the $p_t$ cuts imposed on the
data ensure that the cross section is approximately zero at LO,
such that NLO is effectively LO and NNLO is effectively NLO. Hence the K-factor
calculation is considered unreliable for this lowest bin. For ATLAS the K-factors include correction
from LO to NLO predictions in EW as well as correction from NLO to NNLO in QCD ~\cite{1405.1067}.
All K-factors are close to unity to within $\pm (1-2)\%$, apart from in the low-mass
region $46 \to 66$ GeV when they can reach $5\%$.
Predictions for the ATLAS $W$ and $Z/\gamma^*$ Drell-Yan boson
production are calculated at fixed order in QCD to NNLO and in electroweak to NLO,
using the DYNNLO1.5~\cite{ref2628of1612.03106} program, 
as described in reference~\cite{1612.03016}.
The results are cross-checked with the FEWZ3.1.b2~\cite{ref2425of161203016} program.
The high-mass range is also subject to
background from photon induced di-lepton production which is
estimated using the MRST2004QED~\cite{mrst2004qed} photon PDF.

The QCD evolution equations yield the PDFs
at any value of $Q^2$ if they are parameterised
as functions of $x$ at some initial scale $Q^2_0$. 
In the present analysis, this scale is 
chosen to be $Q^2_0 = 1.9~$\,GeV$^2$ such that it is below the 
charm mass threshold $m_c^2$. 
The heavy quark masses are chosen to be $m_c=1.43~$\,GeV and $m_b=4.5~$\,GeV.
The strong coupling constant is fixed to 
$\alpha_S(M_Z) =  0.118$. A
minimum $Q^2$ cut of $Q^2 \ge 7.5$\,GeV$^2$ is imposed on the HERA data.
This minimum $Q^2$ cut helps to minimise any effects from $ln(1/x)$ resummation~\cite{1710.05935}. These choices follow the ATLASepWZ16 analysis.
The parameter choices are varied to determine the 
systematic uncertainty on fit parameters due to these assumptions.

The quark distributions at the initial scale are represented by the generic form 
\begin{equation}
 xq_i(x) = A_i x^{B_i} (1-x)^{C_i} P_i(x),
\label{eqn:pdf}
\end{equation}
where $P(x)$ is a polynomial in powers of $x$.
The parameterised quark distributions $q_i$ are chosen to be
the valence quark distributions ($xu_v,~xd_v$) and the light anti-quark distributions
($x\bar{u},~x\bar{d},~x\bar{s}$). The gluon distribution 
is parameterised with the more flexible form
$xg(x) = A_g x^{B_g} (1-x)^{C_g}P_g(x)  - A'_g x^{B'_g} (1-x)^{C'_g}$\footnote{In the analysis presented here, $C'_g$ is fixed to $C'_g = 25$.
The exact value of $C'_g$ does not matter provided it is large enough that this negative term does not contribute at large $x$, the present choice follows the HERAPDF~\cite{1506.06042}.}.
The parameters $A_{u_v}$ and $A_{d_v}$ are fixed using 
the quark counting rules and $A_g$ is fixed using the momentum sum rule.
The normalisation and slope parameters, $A$ and $B$,
of $\bar{u}$ and $\bar{d}$  are set equal such that 
$x\bar{u} = x\bar{d}$ at very small $x$. 
The strange PDF 
$x\bar{s}$ is parameterised as in Eq.\,\ref{eqn:pdf}, 
with $P_{\bar{s}}=1$ and $B_{\bar{s}} = B_{\bar{d}}$, leaving 
two free strangeness parameters, $A_s$ and $C_s$.  
By default it is assumed that $xs=x\bar{s}$.
Terms are added in the polynomial
expansion $P_i(x)$ only if  required
by the data, following the procedure described 
in Ref.~\cite{HERA:2009wt}. This leads to
an additional term, $P_{u_v}(x)=1+ E_{u_v} x^2$.
The inclusion of the CMS $Z$ and $W$ data does not alter the preferred form of the
 parameterisation. Thus the form of the central parameterisation follows 
that of the ATLASepWZ16 analysis~\cite{1612.03016}. 
 There are 15 free parameters. However additional parameters,
 which do not yield any significant decrease in $\chi^2$, but which can modify 
the shape of the PDFs, are added as parameterisation variations,  see Section~\ref{sec:results}

The ATLAS data are compared to the theory using the $\chi^2$ function defined
in in Eqs. 21 and 17 of the ATLAS analysis~\cite{1612.03016} which accounts for 
correlated systematic uncertainties of the data using nuisance parameters. 
For the CMS data the correlated systematics are provided as correlation 
and covariance matrices and hence the form of the $\chi^2$ used is as given in 
Eqs. 18 and 19 of the xFitter write-up~\cite{HERAFitter, xFittermanual}.

\section{Results}
\label{sec:results}

The data from CMS and ATLAS are first considered separately.
For both data sets the $W$ production data and the $Z$ mass peak data are 
first considered separately and then together.
The following data subsets are considered, together with the HERA data:
\begin{itemize}
\item {\bf ATLAS W} fit includes the ATLAS 7 TeV pseudo-rapidity distributions of the leptons from 
$W^+$ and $W^-$ decay with full systematic correlations between them;
\item {\bf CMS W7,8} fit includes the CMS asymmetry 
distributions of positive and negative decay muons for 7 TeV data and the 
separate $W^+$ and $W^-$ distribution with full correlations between them for the 8 TeV data,
as a function of muon pseudo-rapidity;
\item {\bf ATLAS Z} fit includes the ATLAS 7\,TeV $Z$ data in the $M_Z$ peak region
$66 < M_Z < 116$\,GeV 
in central, $0 < y < 2.4$ (called CC), and forward, $ 1.2 < y  < 3.6$ (called CF), rapidity regions, 
with full systematic correlations between these rapidity regions;
\item {\bf CMS Z7} fit includes the $Z$-mass peak region is $60 < M_Z < 120$\,GeV
  and the rapidity range is $0 < y < 2.4$;
\item {\bf ATLAS W,Z} fit and {\bf CMS Z7+W7,8} fit include the $W$ and $Z$ data together,
  from ATLAS and CMS, respectively.  
When the ATLAS $W$ and $Z$ data are considered together, the full systematic correlations 
between the ATLAS $W$ and ATLAS $Z$ data are used. There are no equivalent 
correlations for the CMS data.
\end{itemize}
The total $\chi^2$ per number of degrees of freedom (NDF) and the partial 
$\chi^2$ per number of data points (NDP) of the data sets entering the fits 
are listed in Tables~\ref{tab:tab1} and~\ref{tab:tab2} 
for each experiment.
The $\chi^2$ for the CMS 8 TeV $W$ data are small and so a check is made using 
the 8 TeV $W$-asymmetry data. The typical $\chi^2/NDP$ for these data are $5/11$
and the parameters of the fits are essentially unchanged\footnote{
  The CMS PDF fit to these $W$-asymmetry data together with the HERA data gave
  partial $\chi^2$/NDP = 3/11\cite{1603.01803}.
}. 

Then the ATLAS and CMS data are considered together as summarised in 
Table~\ref{tab:tab3}. First the ATLAS and CMS $W$ data are fitted 
together, then the ATLAS and CMS $Z$ data are fitted together and finally 
all the ATLAS and CMS $W$ and $Z$ 
data are considered together. The latter is our main fit and it is referred to
as the CSKK fit hereafter. 

All of the fits describe the HERA data as well as when the HERA data are fitted alone, 
there is no tension between the LHC and HERA data. There is also no tension between 
subsets of the CMS data or subsets of the ATLAS data. Finally 
there is also no strong tension between the ATLAS and the CMS measurements.
 The $\chi^2$ for the CMS data is still good when the ATLAS data are 
 added and conversely the $\chi^2$ for the ATLAS data remain good when the CMS data are included.
\begin{table}
\begin{center}
\begin{tabular}{|c|ccc|}
\hline
\hline
        & CMS Z7          &       CMS W7,8             &CMS Z7 + W7,8  \\
\hline
Total $\chi^2/\rm{NDF}$                          & 1218/1965 & 1225/1074  &  1236/1098 \\
Data set, $\chi^2/\rm{NDP}$           &          &                   & \\
\hline                                             
HERA                & 1156/1056 & 1157/1056  &  1157/1056   \\
CMS 7\,TeV central $Z$     & 11/24     &            &  11/24       \\
CMS 7\,TeV W-asymmetry    &           & 13/11      &  13/11       \\
CMS 8\,TeV $W^+,W^-$    &           & 4/22     &  4/22      \\
\hline
\hline  
\end{tabular}
\end{center}
\caption{\large Total and partial $\chi^2$ for data sets entering the PDF fits of the CMS data.}
\label{tab:tab1}
\end{table}
\begin{table}
\begin{center}
\begin{tabular}{|c|ccc|}
\hline
\hline
             & ATLAS Z &      ATLAS W            & ATLAS W,Z  \\
\hline
Total $\chi^2/\rm{NDF}$               & 1233/1062 & 1245/1063  & 1276/1084 \\
  Data set, $\chi^2/\rm{NDP}$           & &   &  \\
\hline                
HERA                 & 1155/1056 & 1160/1056  & 1164/1056 \\
ATLAS $W^+$           &           & 12/11      & 12/11     \\
ATLAS $W^-$           &           & 8/11     & 9/11    \\
ATLAS $Z$ central CC  & 14/12     &            & 15/12     \\
ATLAS $Z$ central CF  & 9/9     &            & 8/9     \\
\hline
\hline  
\end{tabular}
\end{center}
\caption{\large Total and partial $\chi^2$ for data sets entering the PDF fits of the ATLAS data.}
\label{tab:tab2}
\end{table}
\begin{table}
\begin{center}
\begin{tabular}{|c||c|c|c|}
\hline
\hline
& ATLAS and CMS $W$ & ATLAS and CMS $Z$ & ATLAS and CMS\\
         &  & &$W$ and $Z$, CSKK fit\\
\hline
Total $\chi^2/\rm{NDF}$ & 1265/1096 = 1.15    & 1244/1086 = 1.15 & 1308/1141 = 1.15  \\
 Data set, $\chi^2/\rm{NDP}$          &  & &\\
\hline
HERA                       & 1159/1056  & 1157/1056 & 1163/1056       \\
ATLAS $W^+$              & 12/11      &           & 13/11           \\
ATLAS $W^-$          & 8/11     &           & 9/11            \\
ATLAS central CC $Z$     &            & 14/12     & 16/12           \\
ATLAS central CF $Z$         &            & 9/9     & 7/9             \\
CMS 7\,TeV central $Z$       &            & 12/24     & 12/24           \\
CMS 7\,TeV W-asym.           &    13/11   &           & 14/11           \\
CMS 8\,TeV $W^+,W^-$         &    6/22  &           & 5/22          \\
\hline
\hline  
\end{tabular}
\end{center}
\caption{Total and partial $\chi^2$ for data sets entering the PDF fits of the ATLAS and CMS data.}
\label{tab:tab3}
\end{table}

\begin{figure}[tbp]
  \vspace{-5.5cm}
  \centerline{
    \includegraphics[width=1.1\textwidth]{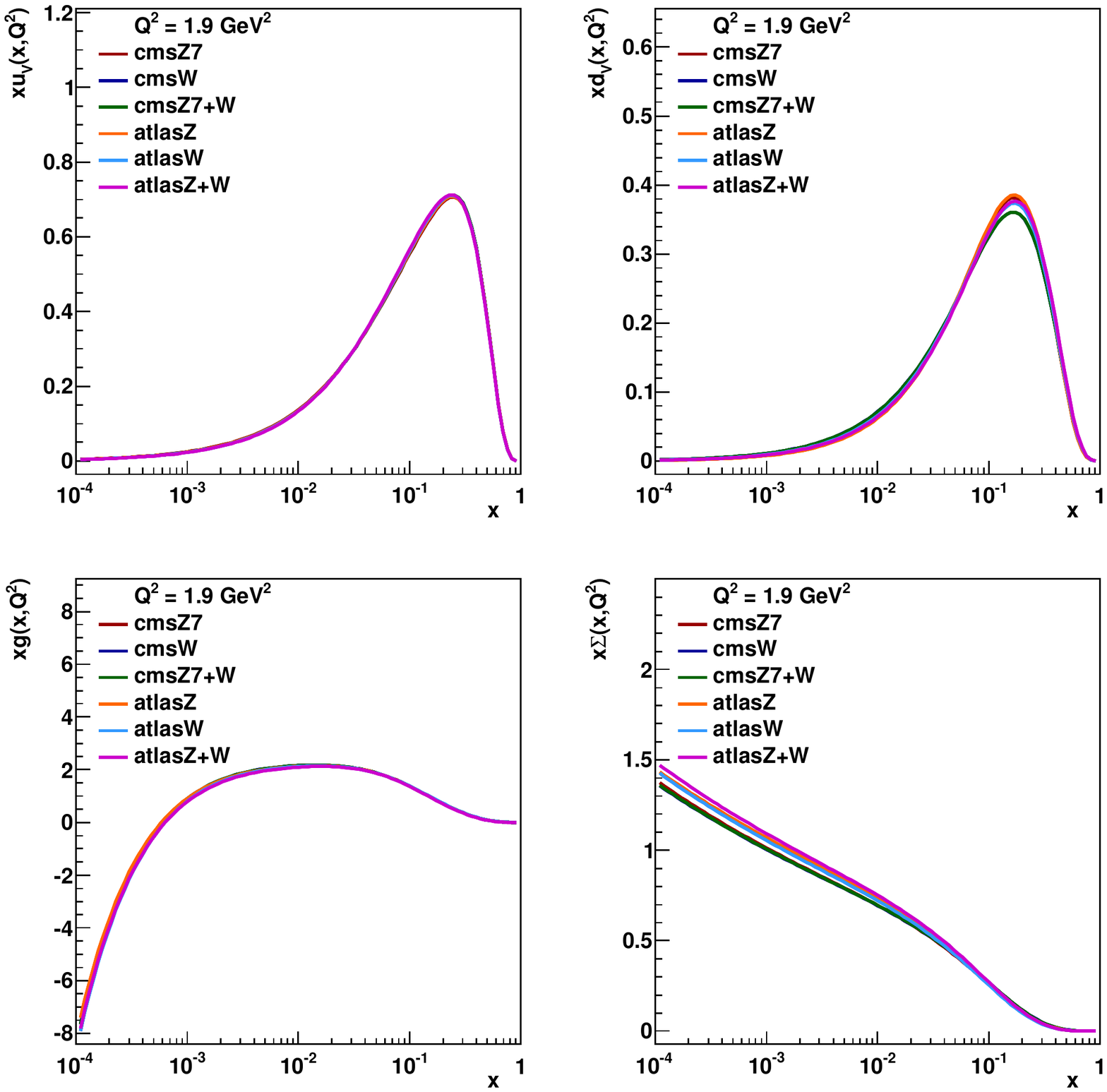}}
  \vspace{-2.5cm}
  \caption {\large The PDFs $u_v$, $d_v$, gluon and total sea, as determined by fits to HERA data plus various subsets of ATLAS and CMS $W$ and $Z$ data. 
Full details are given in the text.
  }
  \label{fig:fig1}
\end{figure}

\begin{figure}[tbp]
  \vspace{-5.5cm}
  \centerline{
    \includegraphics[width=1.1\textwidth]{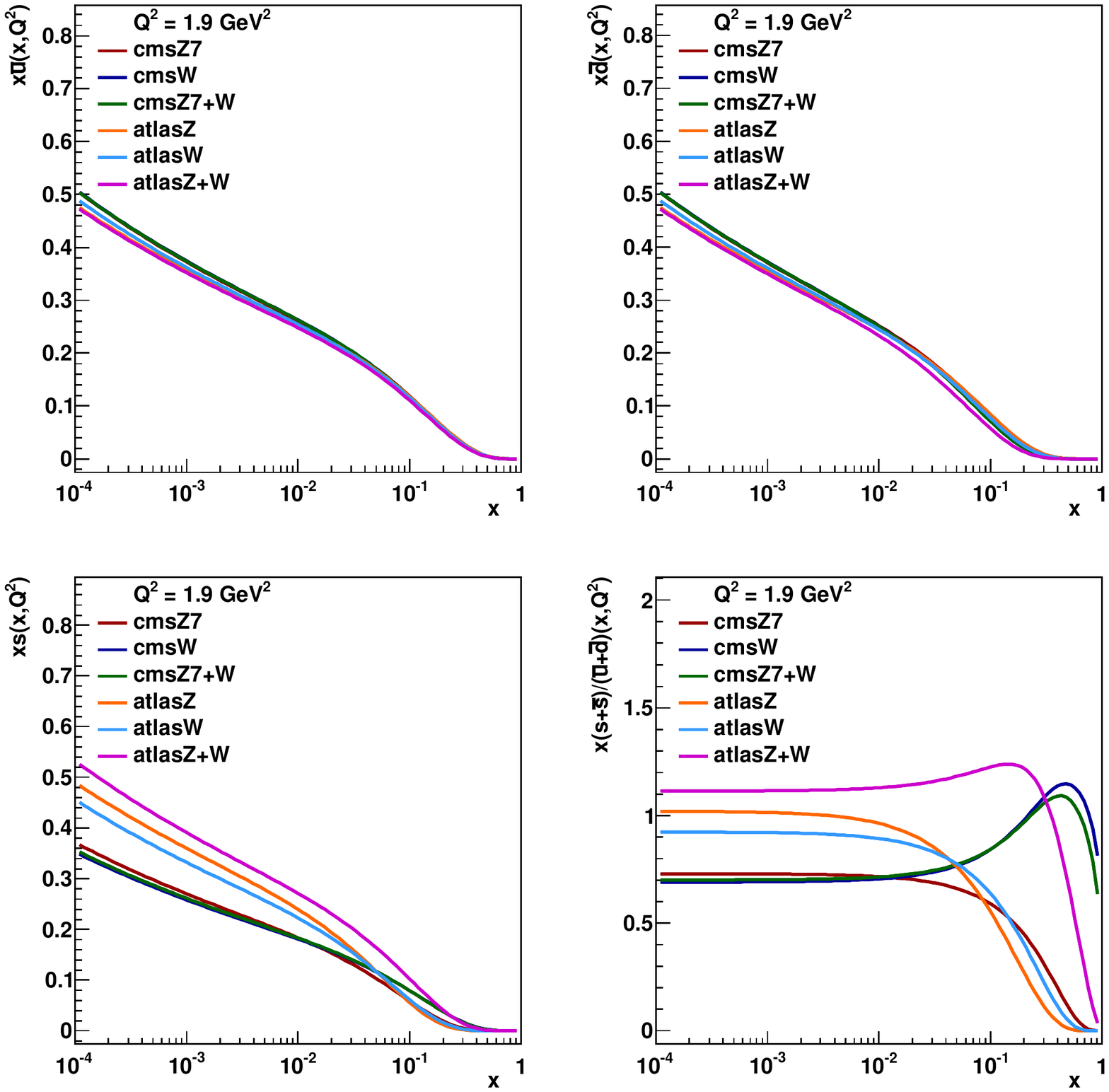}}
  \vspace{-2.5cm}
  \caption {\large  The PDFs $\bar{u}$, $\bar{d}$, $s$ and 
    the ratio $(s +\bar{s})/(\bar{u} +\bar{d}) $ for fits to HERA data plus various subsets
    of ATLAS and CMS $W$ and $Z$ data.  
Full details are given in the text.
  }
  \label{fig:fig2}
\end{figure}

In Figures~\ref{fig:fig1} and~\ref{fig:fig2} the central values of the 
PDFs for the fits to the 
ATLAS and CMS data separately are shown. 
The PDFs illustrated are valence quarks, $u_v$, $d_v$, gluon, $g$, total sea, $\Sigma$, 
antiquarks, $\bar{u}$, $\bar{d}$, $s$ quark and 
the ratio $(s +\bar{s})/(\bar{u} +\bar{d})$. This ratio is unity if strange quarks are not 
suppressed in relation to light quarks and is $\sim 0.5$ for the conventional level of 
suppression.  
 
Figure~\ref{fig:fig1} shows that the valence quarks, gluon and total sea are rather 
similar for all fits--although the shape of $d_v$ differs a little at its peak.
The shapes of these PDFs are well determined by the HERA data.
Figure~\ref{fig:fig2} shows that the flavour break up of the sea, 
which is sensitive to the LHC data, is different between CMS and ATLAS. 
The ATLAS data have larger strangeness than the CMS data. However at small $x$ neither 
the CMS nor the ATLAS data support the conventional level of suppression. 
 The distributions 
differ more markedly for $x > 0.1$ but parameterisation uncertainties become large in 
this region, see later. The ATLAS $W$ and $Z$ data taken separately imply smaller 
strangeness than the ATLAS $W, Z$ data taken together, particularly for $x > 0.1$, but 
there is additional information in the fit to these data coming from the common 
correlated systematic uncertainties.

\begin{figure}[tbp]
  \vspace{-5.5cm}
  \centerline{
    \includegraphics[width=1.1\textwidth]{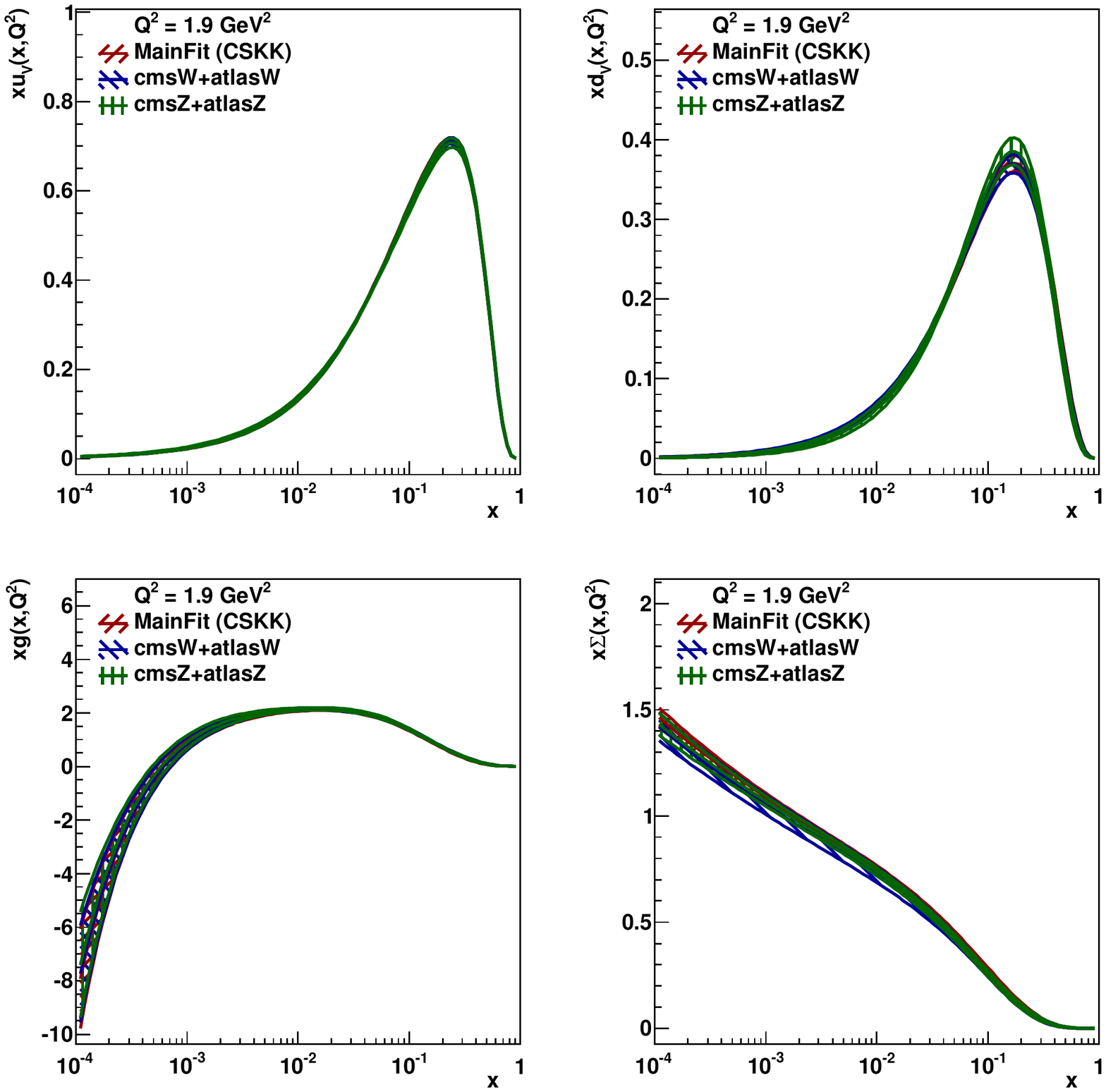}}
  \vspace{-2.5cm}
  \caption {\large The PDFs $u_v$, $d_v$, gluon and total sea, for fits to HERA data plus: 
ATLAS and CMS $W$ data; ATLAS and CMS $Z$ data; and ATLAS and CMS $W$ and $Z$ data together (CSKK).
The bands represent experimental uncertainties. Full details are given in the text.
  }
  \label{fig:fig3}
\end{figure}

\begin{figure}[tbp]
  \vspace{-5.5cm}
  \centerline{
    \includegraphics[width=1.1\textwidth]{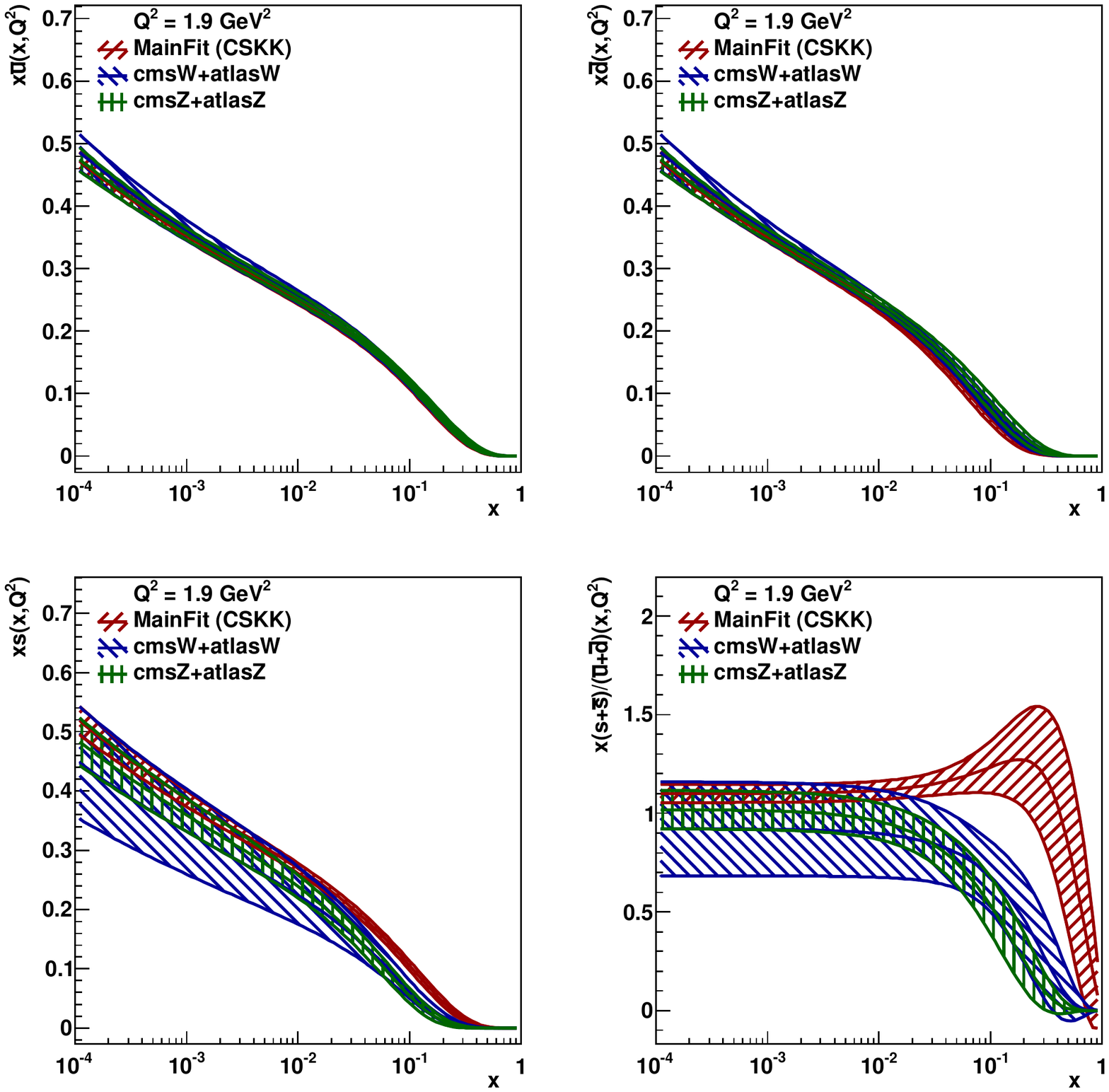}}
  \vspace{-2.5cm}
  \caption {\large  The PDFs $\bar{u}$, $\bar{d}$, $s$ and 
the ratio $(s +\bar{s})/(\bar{u} +\bar{d}) $ for fits to HERA data plus: 
ATLAS and CMS $W$ data; ATLAS and CMS $Z$ data; and ATLAS and CMS $W$ and $Z$ data together (CSKK).
The bands represent experimental uncertainties. Full details are given in the text.
  }
  \label{fig:fig4}
\end{figure}

Figure~\ref{fig:fig3} shows that the valence quarks, gluon and total sea are also  
similar for fits done to the $W$ and $Z$ data separately, and that both are similar to the 
the combination of the $W$ and $Z$ data.
Figure~\ref{fig:fig4} shows that the flavour break up of the sea, 
which is sensitive to the LHC data, is similar at small $x$ for the $W$ and $Z$ data 
taken separately and both data sets support unsuppressed strangeness. 
The PDFs are now shown with their
experimental uncertainties resulting from the 
statistical and correlated and uncorrelated systematic uncertainties on the $ep$ and 
$W,Z$ cross section measurements. 
From Fig.~\ref{fig:fig4} it can also be seen that the $Z$ data are more 
sensitive to strangeness than the $W$ data.
 The distributions 
differ more markedly for $x > 0.1$ where parameterisation uncertainties become large. 
As before, the $W$ and $Z$ data taken separately imply smaller 
strangeness than when $W, Z$ data are taken together, particularly for $x > 0.1$, 
because of the additional information coming from the
correlated systematic uncertainties of the ATLAS data.

\begin{figure}[tbp]
  \vspace{-5.5cm}
  \centerline{
    \includegraphics[width=1.1\textwidth]{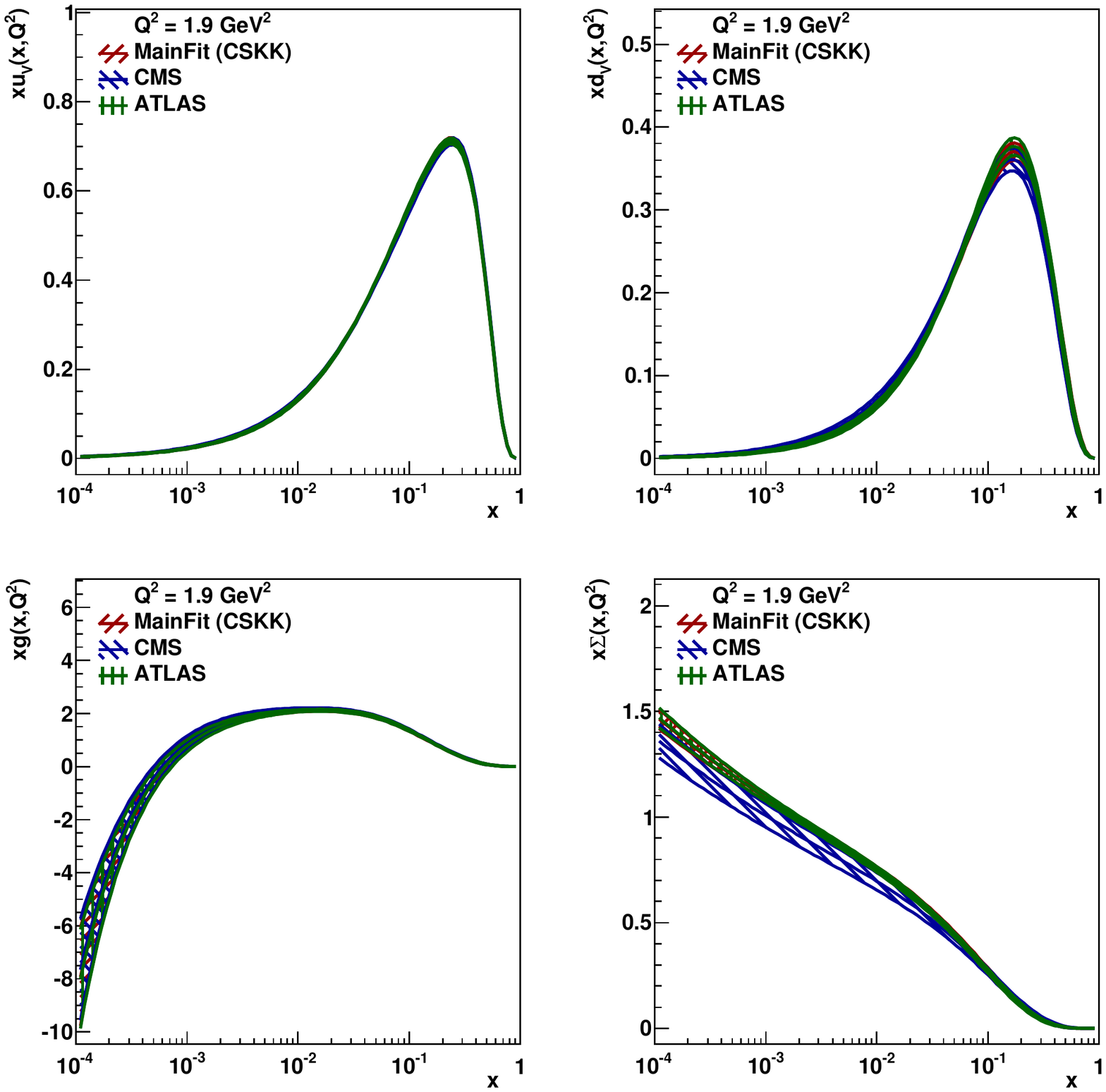}}
    \vspace{-2.5cm}
  \caption {\large The PDFs $u_v$, $d_v$, gluon and total sea, for fits to HERA data plus: 
ATLAS $W$ and $Z$ data; CMS $W$ and $Z$ data; and ATLAS and CMS $W$ and $Z$ data together (CSKK).
The bands represent experimental uncertainties. Full details are given in the text.
  }
  \label{fig:fig5}
\end{figure}

\begin{figure}[tbp]
  \vspace{-5.5cm}
  \centerline{
    \includegraphics[width=1.1\textwidth]{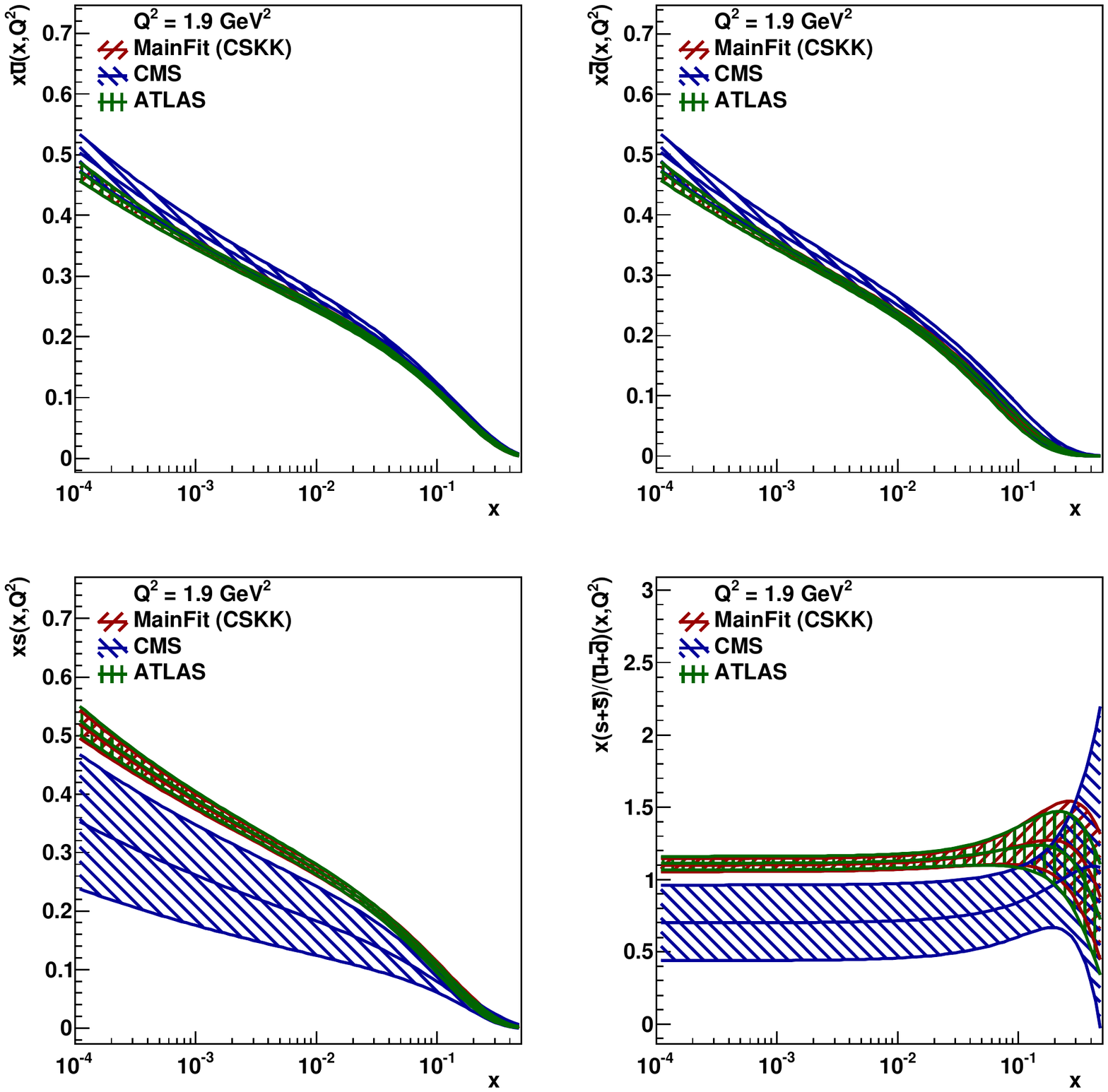}}
  \vspace{-2.5cm}
  \caption {\large  The PDFs $\bar{u}$, $\bar{d}$, $s$ and 
the ratio $(s +\bar{s})/(\bar{u} +\bar{d}) $ for fits to HERA data plus: 
ATLAS $W$ and $Z$ data; CMS $W$ and $Z$ data; and ATLAS and CMS $W$ and $Z$ data together (CSKK).
The bands represent experimental uncertainties. Full details are given in the text.
  }
  \label{fig:fig6}
\end{figure}

Figures~\ref{fig:fig5} and~\ref{fig:fig6} show the PDFs resulting from the fits to the 
CMS $W$ and $Z$ data, the ATLAS $W$ and $Z$ data and the ATLAS and CMS $W$ and $Z$ data. 
These PDFs are also shown with their
experimental uncertainties.  As before,
the PDFs illustrated are $u_v$, $d_v$, gluon and total sea, 
$\bar{u}$, $\bar{d}$, $s$ and 
the ratio $(s +\bar{s})/(\bar{u} +\bar{d})$. The valence quarks, gluon and total sea are 
similar for the PDFs derived from the ATLAS and CMS data, small differences are 
well within the spread of 
uncertainties. Considering the flavour break up one can see that, 
although the strange distributions from ATLAS and CMS differ, 
in the  region of maximal 
sensitivity of the LHC data, $x\sim 0.01$,
the CMS data alone imply a strangeness ratio only $\sim 1-2$ standard 
deviations lower than that of the ATLAS data alone.

Considering the CSKK fit to both ATLAS and CMS data it is clear that
the greater accuracy of the ATLAS data dominates the fit, 
such that the combined fit has unsuppressed strangeness. It is interesting to 
consider the separate effect of the CMS and ATLAS $W$ and $Z$ samples on the PDF uncertainties for
different PDFs. Figure~\ref{fig:fig7} shows the effect on the uncertainties of the 
valence and sea quarks, of adding the four different data samples,
namely the ATLAS $W$, ATLAS $Z$, CMS $W$ and CMS $Z$, separately to the HERA data. 
The effect on valence quarks is illustrated by the $d$-valence. One can see that the $W$ data
of ATLAS and CMS are equally powerful in constraining the valence distributions, whereas the $Z$ 
data do not have such a powerful effect. For the total sea distribution, $\Sigma$, the ATLAS $Z$ 
data gives the most powerful constraint, followed by the ATLAS $W$, CMS $W$ and CMS $Z$ data sets. 
The same ordering is seen in the $\bar{u}$ and $\bar{d}$ uncertainties and is most pronounced for 
$s$ and 
the ratio $(s +\bar{s})/(\bar{u} +\bar{d})$, as illustrated in the bottom two plots.  
 
The CMS data are not in tension with the ATLAS data since 
the $\chi^2$ for the CMS data is still remains very good when ATLAS data are added.
A further measure of the consistency of the data in the fit can be seen by considering 
the shifts of
the systematic uncertainties, as determined by the nuisance parameters in the fit. 
The distribution of the shifts is shown in Fig.~\ref{fig:fig8}, for all the
correlated systematic uncertainties and separately for the ATLAS data alone.

The default method for assessing experimental uncertainties is the Hessian method based on the $\Delta\chi^2=1$,
however we have also use the MC replica method~\cite{mcreplicas} 
which can differ from the Hessian when experimental uncertainties are non-Gaussian.
The PDFs obtained with both methods agree well.
A comparison of the uncertainties on $u$-valence, $d$-valence, 
$\Sigma$ and the ratio $(s +\bar{s})/(\bar{u} +\bar{d})$ for the CSKK fit is shown in 
Fig.~\ref{fig:fig9}. 
The uncertainty estimates of these two methods are compatible.

Figures~\ref{fig:fig10}--\ref{fig:fig14} 
show comparisons of the predictions of the CSKK fit to the
input data sets for both ATLAS and CMS data. 
For the ATLAS data the theoretical prediction is shown both with and 
without the shifts induced  by the nuisance parameter treatment of the
correlated systematic uncertainties.

We now consider adding the off-peak Drell-Yan data at 7 TeV, for both ATLAS and CMS. 
First we add the high-mass data, in the dilepton-mass range $116$ to $136$\,GeV,
 for ATLAS and, in the ranges
$120$ to $200$\,GeV, and $200$ to $1500$\,GeV, for CMS 
(called CSKK+highMass). 
Then we add the low-mass data, 
in the dilepton-mass range $46$ to $66$\,GeV, for ATLAS and in the ranges
$30$ to $45$\,GeV  and $45$ to $60$\,GeV for CMS (called CSKK+highMass+lowMass).
The correlations between these data and the central $Z$-peak data are accounted for in all cases. 
The resulting PDFs are shown, with their experimental uncertainties, in 
Figures~\ref{fig:fig15} and~\ref{fig:fig16} compared to the nominal CSKK fit.
These figures show that adding the low and high-mass 
regions of the ATLAS and CMS data does not change the result substantially. 
The experimental uncertainties are also not much reduced.
The $\chi^2$s for the fit including all the off-peak data are given in 
Table~\ref{tab:tab4}. The off-peak 7 TeV $Z$ data are generally well described.
The high $\chi^2$ for the ATLAS low-mass data was already observed in 
the ATLAS analysis~\cite{1612.03016}. This data set has little influence on the fit.
Figures~\ref{fig:fig17} and~\ref{fig:fig18} show comparisons of 
the off-peak mass bin data to the predictions of the CSKK+highMass+lowMass fit.
Since there are larger theoretical uncertainties for these 
off-peak mass regions~\cite{1612.03016,1710.05935}, coming from electroweak effects and photon induced processes,
we chose to use only the peak data for the nominal CSKK fit. 

\begin{table}
\begin{center}
\begin{tabular}{|c||c|c|c|}
\hline
\hline
&\multicolumn{2}{c|}{ATLAS and CMS $W$ and all $Z$ bins}&\multicolumn{1}{c|}{CMS $W$ and}\\
    &  $Z$ at 7\,TeV     & $Z$ at 7 and 8\,TeV & all $Z$ bins\\
\hline
Total $\chi^2/\rm{NDF}$                 & 1481/1243 = 1.19  & 1814/1351 = 1.34  &  1596/1290 = 1.24  \\
Data set, $\chi^2/\rm{NDP}$    &       & &\\
\hline                                                     
HERA                    & 1163/1056         & 1178/1056  &   1186/1056      \\
ATLAS $W^+$             &  13/11            &  12/11     &       \\
ATLAS $W^-$             &   9/11            &  15/11     &   \\
ATLAS central CC $Z$    &  15/12            &  26/12     &  \\
ATLAS central CF $Z$    &   7/9             &  8/9       &  \\
ATLAS CC $Z$, $116 < M_z < 150$~\GeV       &   8/6             &     7/6   & \\
ATLAS CF $Z$, $116 < M_z < 150$~\GeV      &   4/6             &     4/6    &  \\
ATLAS CC $Z$, $46 < M_z < 66$~\GeV        &  28/6             &      34/6  & \\
CMS 7\,TeV W-asym.       &  14/11            &  14/11  & 18/11 \\
CMS 8\,TeV $W^+, W^-$       &     5/22          &       7/22 & 5/22\\
CMS 7\,TeV $Z$ central   &  12/24            &  13/24        & 16/24\\
CMS 7\,TeV $Z$,  $120 < M_z < 200$~\GeV   &  31/24            &  28/24     &   25/25  \\
CMS 7\,TeV $Z$,  $200 < M_z < 1500$~\GeV  &  20/12            &  19/12     &    17/12 \\
CMS 7\,TeV $Z$,  $30 < M_z < 45$~\GeV   &  35/24            &  35/24       &   36/24\\
CMS 7\,TeV $Z$,  $45 < M_z < 60$~\GeV    &  22/24            &  20/24      &   20/24 \\
CMS 8\,TeV $Z$ central   & &  74/24    &   66/24   \\
CMS 8\,TeV $Z$,  $120 < M_z < 200$~\GeV   & &  73/24      &  56/24  \\
CMS 8\,TeV $Z$,  $200 < M_z < 1500$~\GeV   & &  14/12     &  12/12   \\
CMS 8\,TeV $Z$,   $30 < M_z < 45$~\GeV   & &  38/24       &  37/24 \\
CMS 8\,TeV $Z$,   $45 < M_z < 60$~\GeV   & &  29/24       &  20/24 \\
\hline
\hline  
\end{tabular}
\end{center}
\caption{Total and partial $\chi^2$ for data sets entering the extended PDF fits of 
the ATLAS and CMS data to include off-peak Drell-Yan data. Full details are given in the text}
\label{tab:tab4}
\end{table}

Finally, the impact of the CMS $Z$ double-differential DY data at 8\,TeV are studied.
The invariant mass and rapidity ranges are the same as for the 7\,TeV data.
First just the central $Z$-peak bin is added to the 
CSKK+highMass+lowMass fit, this fit is labelled CSKK+highMass+lowMass+z8Peak.
Then all the 8\,TeV mass bins are added 
(excluding the lowest mass bin $20 < M_Z < 30$ GeV, just as for the 7\, TeV data),
this fit is labelled CSKK+highMass+lowMass+z8All. 
Since the $u_v$, $d_v$, gluon and total sea PDFs remain very similar to those of the nominal CSKK fit,
we choose to shown only the central values of the 
$\bar{u}$, $\bar{d}$, $s$ PDFs and the ratio $(s +\bar{s})/(\bar{u} +\bar{d})$ in 
Figure~\ref{fig:fig19}, compared to the CSKK fit with its experimental uncertainties.
The $\chi^2$ are given in Table~\ref{tab:tab4}.
The 8\,TeV CMS $Z$ data are not well fitted, especially for the central mass region.
This results in an overall significantly larger  $\chi^2$, as also found by 
NNPDF3.1~\cite{nnpdf31}. When these data are fitted together with the ATLAS data some tension 
with the ATLAS central mass, central rapidity $Z$ data appears. However, as also shown in 
Table~\ref{tab:tab4}, these 8\,TeV CMS $Z$ data are not well fitted even when they are 
fitted together with just HERA data and other CMS data. 
The input of these 8\,TeV CMS $Z$ data does not change our results substantially in the region of 
sensitivity of the LHC data $x\sim 0.01$, particularly in the case of the 
the central $Z$ peak data. 

For the final combination of the ATLAS and CMS data, model and 
parameterisation uncertainties are considered.
The model uncertainty includes variation of the minimum $Q^2$ cut value, 
the value of the starting scale and the heavy quark masses, following the 
HERAPDF2.0 analysis~\cite{1506.06042}. The variation of the charm quark mass and the 
starting scale are performed simultaneously as the constraint $Q^2_0< m^2_c$ has to be 
fulfilled. The variations are summarised in Table~\ref{tab:tab5} together with
the corresponding values of the total $\chi^2/\rm{NDF}$ and the ratio
$R_s = \frac{s+\bar{s}}{\bar{d}+\bar{u}}$ at $x=0.023$ and $Q^2_0=1.9$\;GeV$^2$, and at $x=0.013$ 
and $Q^2 = M_Z^2$. The latter $x, Q^2$ point represents the region of maximal 
sensitivity of the LHC data and the former $x, Q^2$ point is the corresponding point 
at the starting scale for QCD evolution in the PDF fit.

\begin{table}[tbp]
\renewcommand*{\arraystretch}{1.6}
\begin{center}
\begin{tabular}{|lccc|}
  \hline
  \hline
\multicolumn{1}{|c}{Variation} &
\multicolumn{1}{c}{Total $\chi^2/$NDF} &
\multicolumn{2}{c|}{$R_s = \frac{s+\bar{s}}{\bar{d}+\bar{u}}$} \\
\multicolumn{1}{|c}{} &
\multicolumn{1}{c}{} &
\multicolumn{1}{c}{$x=0.023$,} &
\multicolumn{1}{c|}{$x=0.013$,} \\
\multicolumn{1}{|c}{} &
\multicolumn{1}{c}{} &
\multicolumn{1}{c}{\;\;$Q^2_0=1.9$\;GeV$^2$} &
\multicolumn{1}{c|}{$Q^2_0=8317$\;GeV$^2$} \\
\hline
Nominal CSKK fit & 1308 / 1141 & 1.14  & 1.05\\
\hline
\multicolumn{4}{|c|}{Model variations}\\
\hline
$Q^2_{\rm min} = 5$\;GeV$^2$ & 1375 / 1188  &  1.14  & 1.06  \\
$Q^2_{\rm min} = 10$\;GeV$^2$ & 1251 / 1101 &  1.14  & 1.05  \\
$m_b = 4.25$\;GeV    &  1307 / 1141 &  1.12& 1.04  \\
$m_b = 4.75$\;GeV    &  1310 / 1141 &  1.16& 1.06  \\
$\mu_{f_{0}}^2 = 1.6$\;GeV$^2$ and $m_c = 1.37$\;GeV&  1312 / 1141   &  1.16& 1.06  \\
$\mu_{f_{0}}^2 = 2.2$\;GeV$^2$ and $m_c = 1.49$\;GeV & 1308 / 1141     & 1.12 & 1.05  \\
\hline
\multicolumn{4}{|c|}{Parameterisation variations}   \\
\hline
$B_{\bar{s}}$ &  1308 / 1140     &  1.12 & 1.05  \\
$D_{u_v}$ &   1308 / 1140    & 1.13  & 1.05  \\
$D_{d_v}$ &    1308 / 1140    &  1.14 & 1.05  \\
$D_{g}$ &    1306 / 1140   & 1.15  & 1.06  \\
$D_{\bar{u}}$ &  1305 / 1140     & 1.15  & 1.06  \\
$D_{\bar{d}}$ &  1302 / 1140     & 1.09  & 1.04  \\
$E_{d_v}$ &     1308 / 1140  & 1.14  & 1.05  \\
$A_{\bar{u}}$ and $B_{\bar{u}}$ free &  1306 / 1139     & 1.17  & 1.07  \\
$A_{\bar{u}}$ and $B_{\bar{u}}$ and $B_{\bar{s}}$ free &  1306 / 1138     & 1.17  & 1.07  \\
\hline
\multicolumn{4}{|c|}{$\alpha_s(M_Z)$ variations}   \\
\hline
$\alpha_s(M_Z) = 0.116 $         & 1308 / 1141 & 1.12&  1.04 \\
$\alpha_s(M_Z) = 0.117 $         & 1308 / 1141 & 1.13&  1.05 \\
$\alpha_s(M_Z) = 0.119 $         & 1309 / 1141 & 1.14&  1.06 \\
$\alpha_s(M_Z) = 0.120 $         & 1310 / 1141 & 1.15&  1.06 \\
\hline
\hline
\end{tabular}
\end{center}
\caption{Overview of the impact of variations in the QCD fit of model input parameters, parameterisation and the value of $\alpha_s(M_Z)$, as compared to the nominal fit.  For each variation the total fit
 $\chi^2/$NDF is given as well as the
  values of the quantity $R_s = \frac{s+\bar{s}}{\bar{d}+\bar{u}}$, at $x=0.023$ and $Q^2_0=1.9$\;GeV$^2$ and at $x=0.013$ and $Q^2_0=8317$\;GeV$^2$.
 }
\label{tab:tab5}
\end{table}

The parameterisation uncertainty corresponds to the envelope of the results
obtained by adding extra parameters which can result in somewhat different parton distributions with almost equally good
$\chi^2$ as for the nominal fit. Just as in the ATLASepWZ16 analysis, the parameterisation 
variations considered
 consist of adding extra terms in the polynomials 
$P$ of Eq.\,\ref{eqn:pdf} and allowing the low-$x$ parameter for the strange quark
$B_{\bar{s}}$ to be free. However in addition, further variations of the low-$x$ sea 
quark parameterisation are considered. 
Firstly, the constraint $(x\bar{u} - x\bar{d}) \rightarrow 0$
for $x \rightarrow 0$ is relaxed by allowing the $A$ and $B$ parameters for $\bar{u}$ and $\bar{d}$ to differ. 
The low-$x$ $x\bar{d}(x)$ distribution is found to be consistent with 
$x\bar{u}(x)$,
within experimental uncertainties and the strangeness ratio is still 
consistent with unity, as shown in Figure~\ref{fig:fig20}.
Secondly, the $B_{\bar{s}}$ parameter was allowed to vary in the fit, additionally to allowing the $A$ and $B$
parameters for $\bar{u}$ and $\bar{d}$ to differ.
The low-$x$ $x\bar{d}(x)$ distribution is still found to be consistent with 
$x\bar{u}(x)$, and the low-$x$ strange distribution is somewhat enhanced, similarly to when $B_{\bar{s}}$ is allowed
to be free but the constraint  $(x\bar{u} - x\bar{d}) \rightarrow 0$
for $x \rightarrow 0$ is preserved.
Figure~\ref{fig:fig20} shows a comparison of the sea PDFs from these two fits to those of the nominal CSKK fit.
The valence and gluon PDFs do not differ significantly. These parameterisation variations are also 
summarised in Table~\ref{tab:tab5} together with
the corresponding values of the total $\chi^2/\rm{NDF}$ and the $R_s$ ratio
at $x=0.023$ and $Q^2_0=1.9$\;GeV$^2$ and at $x=0.013$ and $Q^2 = M_Z^2$.

The $\alpha_s$ uncertainty corresponds to a variation of $\alpha_s(M_Z)$ from $0.116$ to $0.120$.
Further uncertainties due to scale choice or the limitations of NNLO calculations are beyond the
scope of the present study. They have been considered in the ATLAS analysis~\cite{1612.03016}.

Figure~\ref{fig:fig21} shows the strangeness ratio resulting from the nominal CSKK fit with
the model and parameterisation uncertainties as well as with the experimental uncertainties.
The ratio is shown as a function of $x$ at the starting scale $Q^2_0$ and at $Q^2=M^2_Z$.
The total uncertainty is dominated by the parameterisation uncertainty for most of the $x$ range.  
The  ratio of the strange to the light quark sea density is found to be consistent with unity at low $x$.
These plots constitute our main result on the strangeness fraction as a function of $x$.
However for direct comparison to the result of the ATLASepWZ16 
analysis~\cite{1612.03016} and NNPDF3.1~\cite{nnpdf31}, we also quote the value of 
strangeness ratio $R_s$ at $x=0.023$ and $Q^2_0=1.9$\;GeV$^2$  
\begin{equation}
 R_s = 1.14 \pm 0.05 \rm {~(experimental)} \pm 0.03 \rm{ ~(model)~} ^{+0.03}_{-0.05} \rm{~(parameterisation)} ~^{+0.01}_{-0.02} \rm{ ~(\alpha_s)}
  \end{equation}
and at $x=0.013$ and $Q^2 = M_Z^2$
\begin{equation}
 R_s = 1.05 \pm 0.02 \rm {~(experimental)~} ^{+0.02}_{-0.01} \rm{ ~(model)~} ^{+0.02}_{-0.01} \rm{~(parameterisation)}
 \pm 0.01 \rm{ ~(\alpha_s)}.
  \end{equation}
The largest uncertainties come from freeing the strange sea $B_{\bar{s}}$~parameter and the additional parameters for $\bar{u}$ and $\bar{d}$. 

One further cross check of the robustness of the present result is performed. 
For the CSKK fit, $\bar{d}-\bar{u}$ at $x\sim 0.1$ is negative, 
approximately 2-3 standard deviations away from the positive value suggested by
the E866 fixed-target Drell-Yan data~\cite{e866}. 
It has been suggested that if positive ($\bar{d}-\bar{u}$)  were 
imposed on the fit~\cite{abm}, the strangeness would decrease since larger 
$\bar{d}$ is correlated to
smaller strangeness in the current parameterisation. However the E866 observation is made at $x \sim 0.1$, 
whereas the LHC data have largest constraining power at $x \sim 0.01$, such that a strong correlation 
between $\bar{d}$ and strangeness at different $x$ values is avoided by allowing differing $x$ shapes
in the parameterisation. 
As a cross-check a fit was made with a parameterisation which forces ($\bar{d}-\bar{u}$) to be in agreement
with the E866 data~\cite{zeusfitter}. The resulting strangeness ratio 
is $R_s = 0.95\pm 0.07 \rm {~(experimental)}$ at $x=0.023$ and $Q^2_0 = 1.9$\;GeV$^2$.
This still represents a 
strangeness ratio which is consistent with unity, although it lies $\sim 2$ standard 
deviations lower than our central result. It is not included in the parameterisation 
variations since it does not represent a good fit to the HERA, 
ATLAS and CMS data.
 The $\chi^2/$NDF of this fit is 1363/1141 for these data 
compared to 1308/1141 for the nominal CSKK fit.

\section{Summary}

The ATLAS and CMS inclusive $W$ and $Z$ differential cross section data at 7 and 8 TeV have 
been analysed within the framework of an NNLO pQCD PDF fit together with the combined HERA 
inclusive cross section data. There is no tension between the HERA data and the LHC data, 
or between the LHC data sets.
The key observation of the present paper is that the LHC data support unsuppressed 
strangeness in the proton
at low $x$ at both low and high scales. 
The result is dominated by the ATLAS data but is not in 
contradiction with the CMS data. 

\section{Acknowledgements}

We would like to thank Juan Rojo for supplying the APPLGRID cross section calculations
for the 7 and 8 TeV CMS $Z$ data
and Lucian Harland-Lang and Robert Thorne for supplying NNLO/NLO K-factors for these data.
We thank our colleagues in CMS and ATLAS who were involved in the original data analyses and our
xFitter colleagues for the development of the public tools  which made this analysis possible.


\clearpage

\clearpage

\clearpage

\clearpage

\clearpage

\clearpage

\clearpage

\begin{figure}[tbp]
  \vspace{-3.5cm}
  \centerline{
    \includegraphics[width=.55\textwidth]{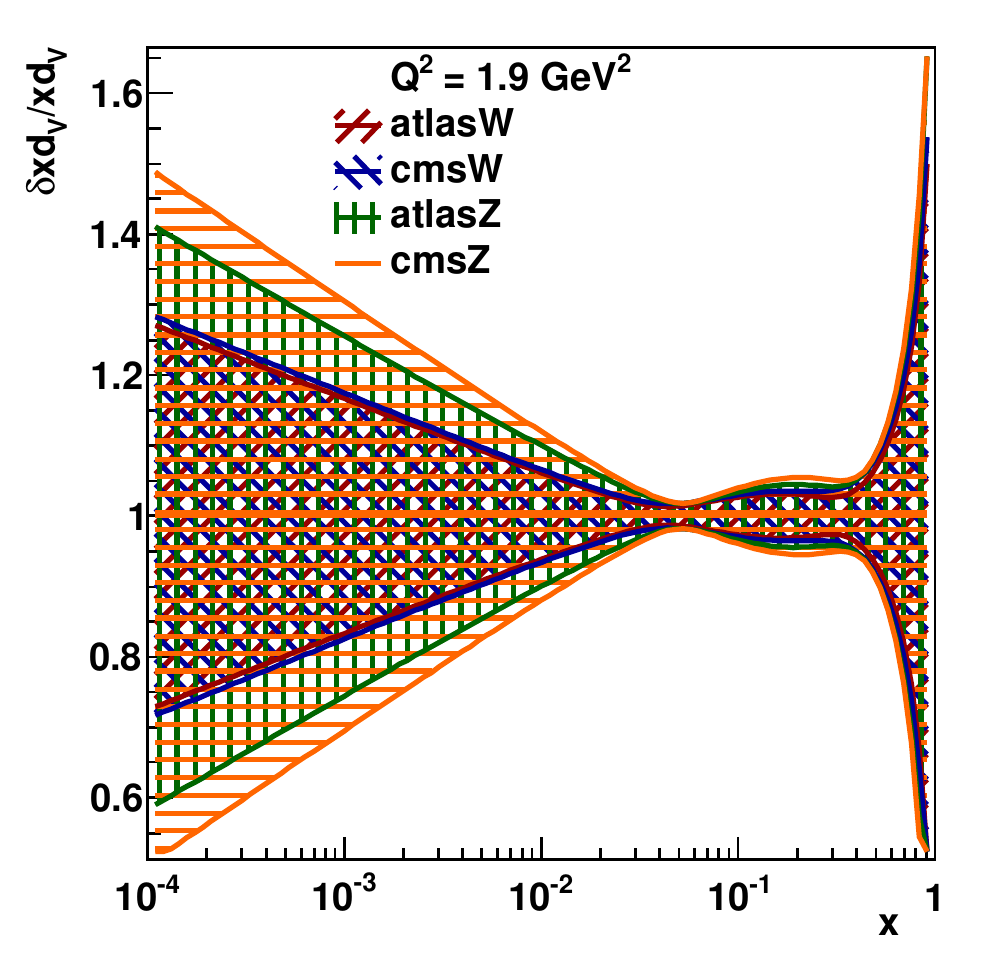}
    \includegraphics[width=.55\textwidth]{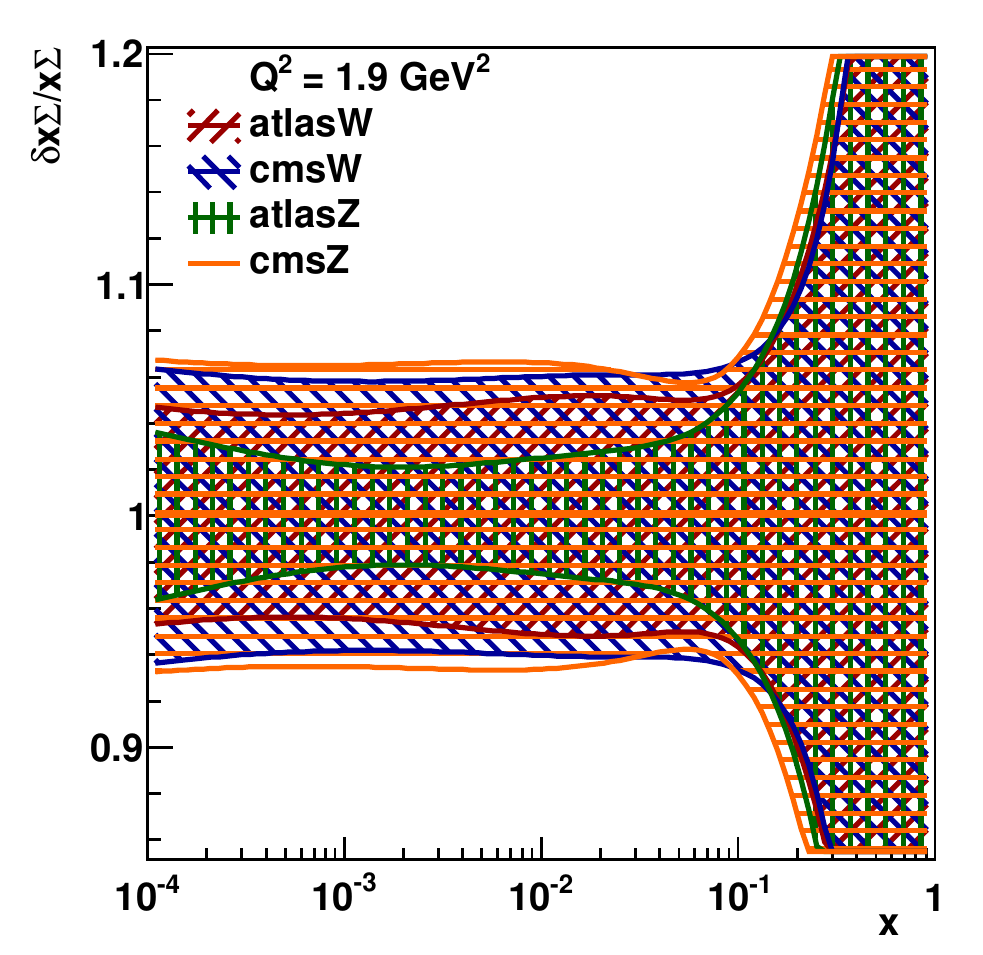}}
\end{figure}
\begin{figure}[tbp]
  \vspace{-7.5cm}
  \centerline{
  \includegraphics[width=.55\textwidth]{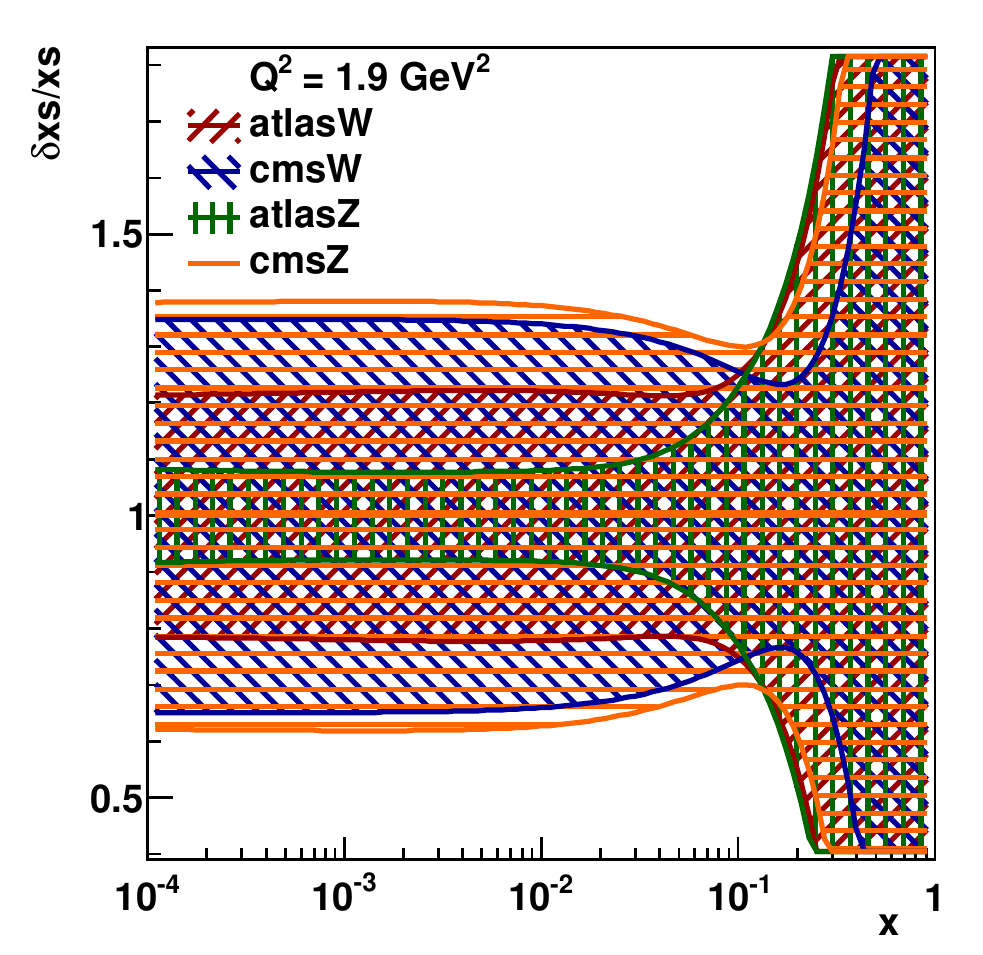}
  \includegraphics[width=.55\textwidth]{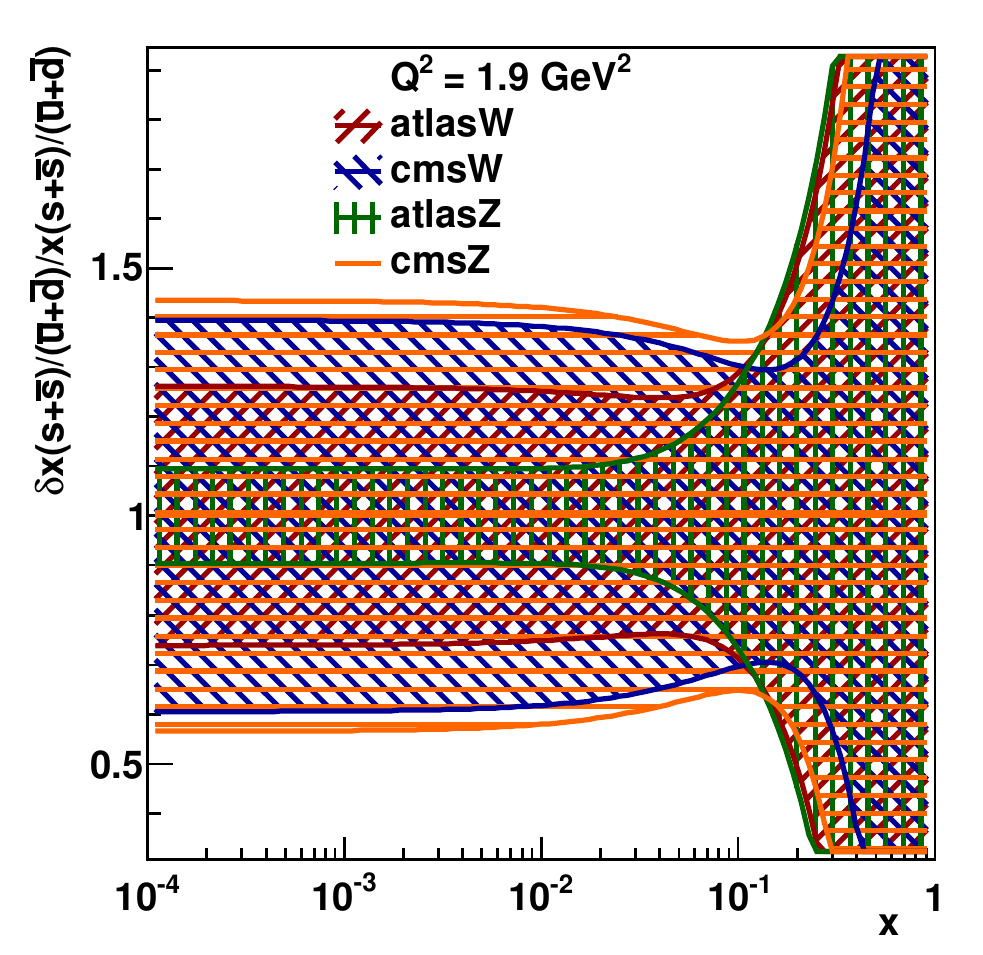}}
  \vspace{.5cm}
  \caption {\large PDF uncertainties for the $d$ valence, total Sea $\Sigma$, $s$ and 
the ratio $(s +\bar{s})/(\bar{u} +\bar{d}) $ for fits to HERA data plus: 
ATLAS $W$; CMS $W$; ATLAS $Z$; CMS $Z$.
The bands represent experimental uncertainties. Full details are given in the text.
  }
  \label{fig:fig7}
\end{figure}

\clearpage

\begin{figure}[tbp]
  \vspace{-5.5cm}
  \centerline{
    \includegraphics[width=1.\textwidth]{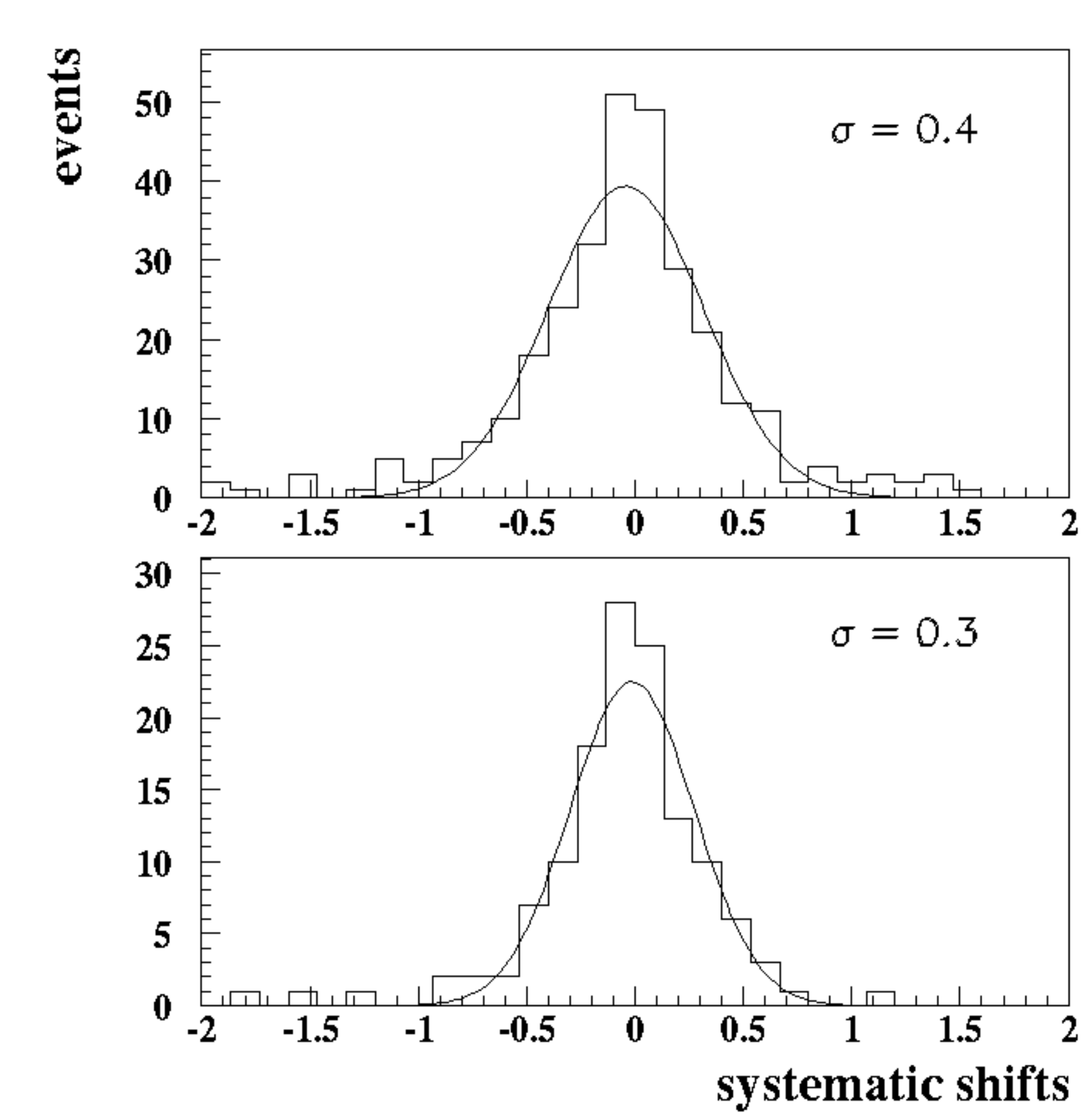}}
  \caption {\large  Distribution of shifts of the correlated systematic uncertainties
    including global normalisations for all uncertainties (top) and for the ATLAS uncertainties only (bottom).
    There are no entries outside the histogram range.
    Gaussian fits to the distributions are shown.
  }
  \label{fig:fig8}
\end{figure}

\clearpage

\begin{figure}[tbp]
  \vspace{-3.5cm}
  \centerline{
    \includegraphics[width=.55\textwidth]{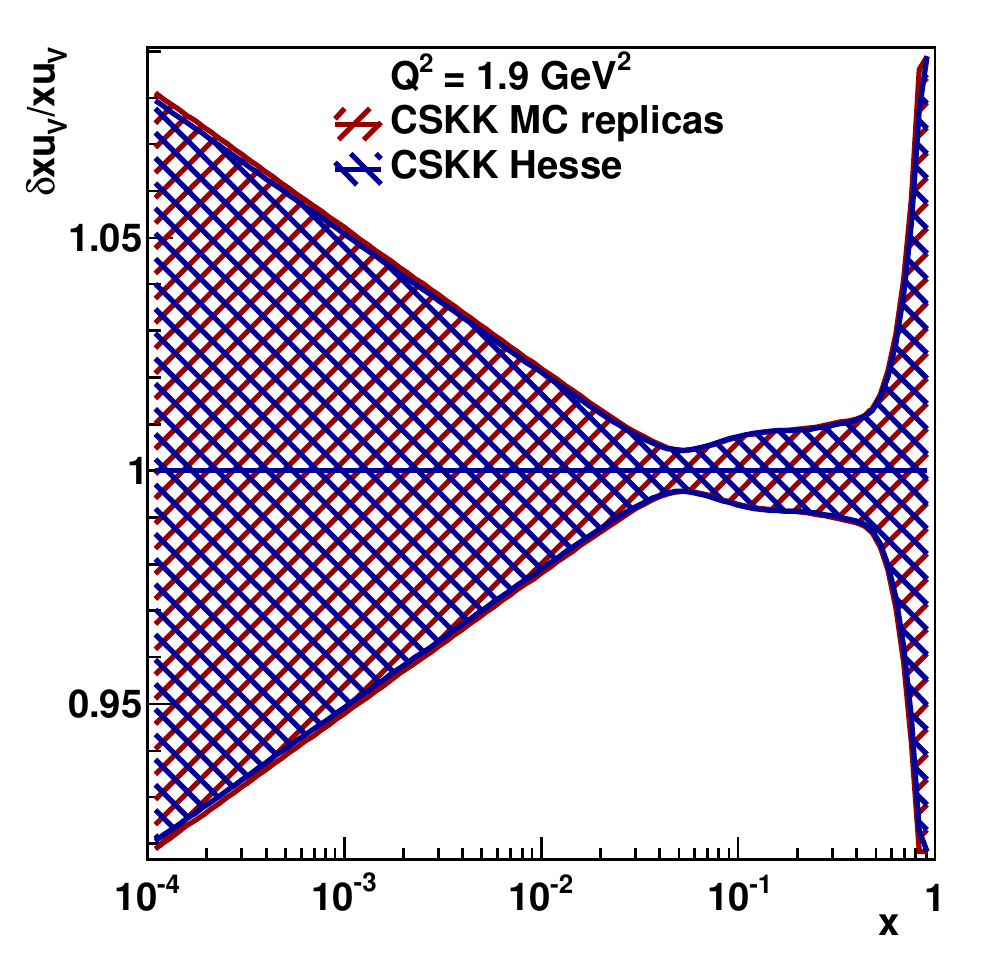}
    \includegraphics[width=.55\textwidth]{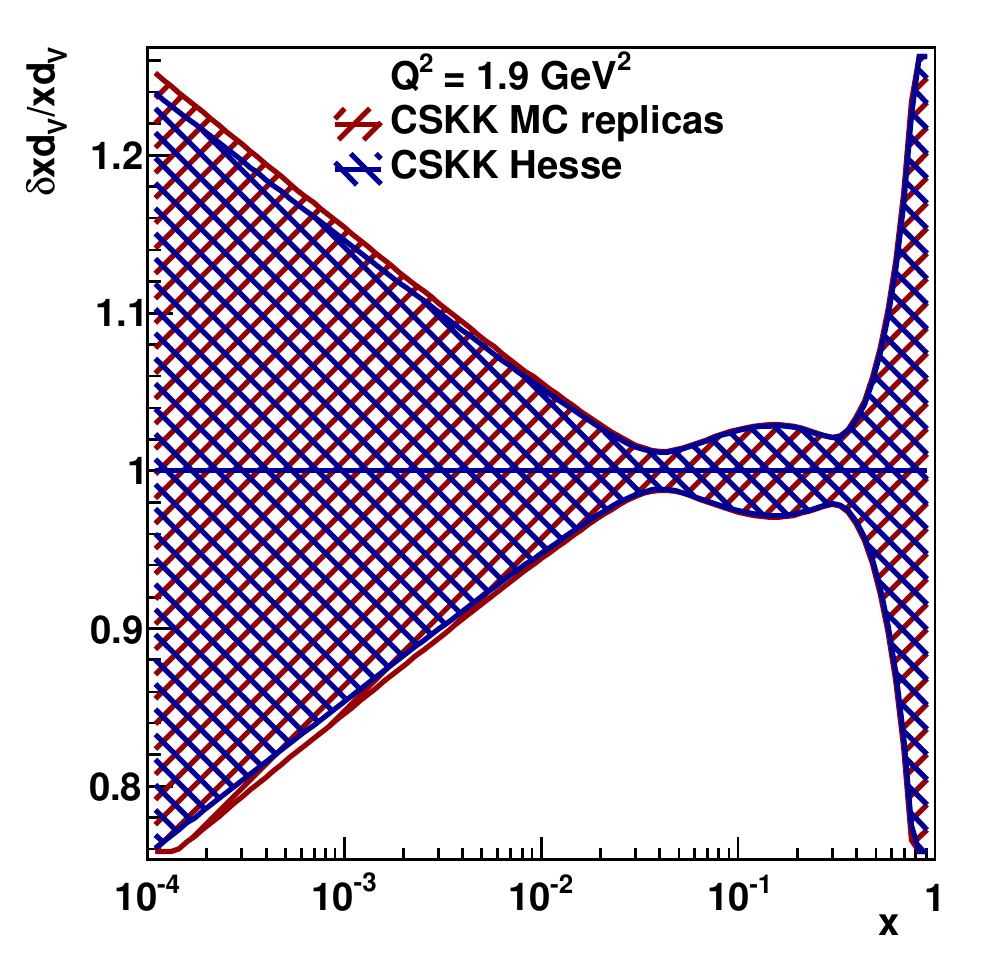}}
\end{figure}
\begin{figure}[tbp]
  \vspace{-7.5cm}
  \centerline{
  \includegraphics[width=.55\textwidth]{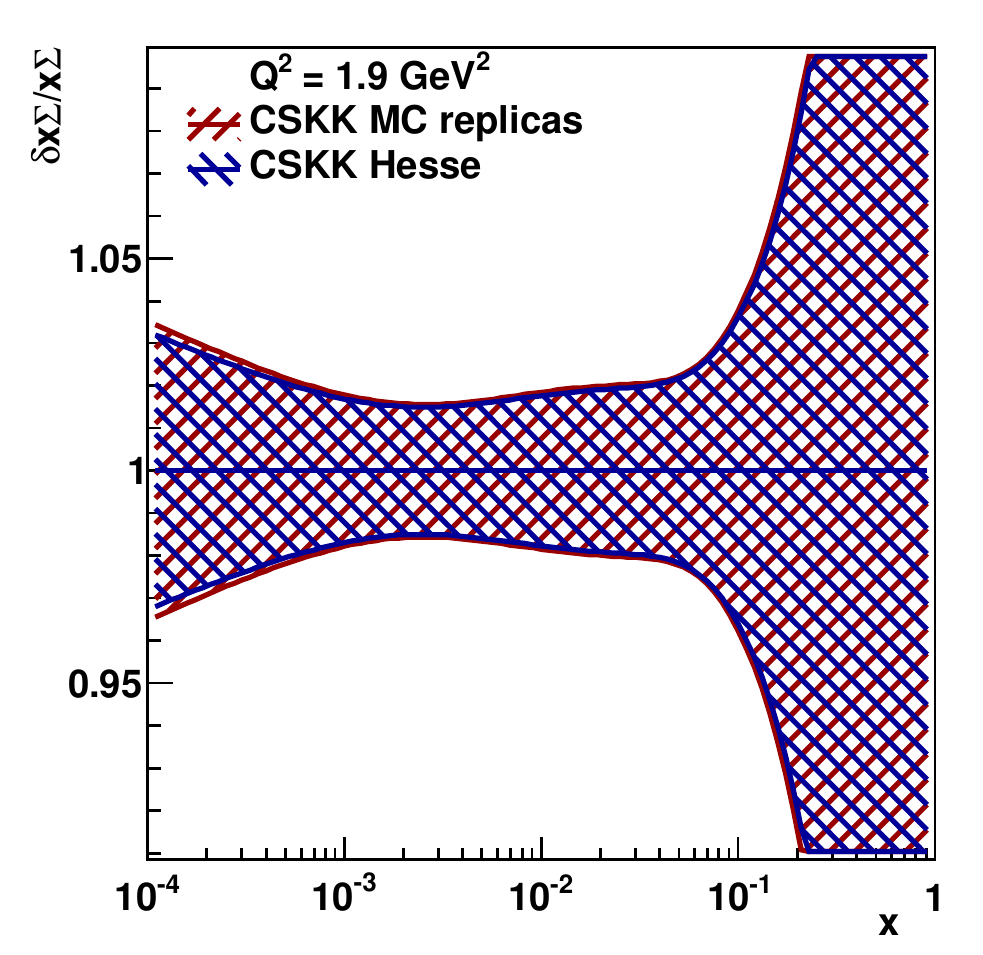}
  \includegraphics[width=.55\textwidth]{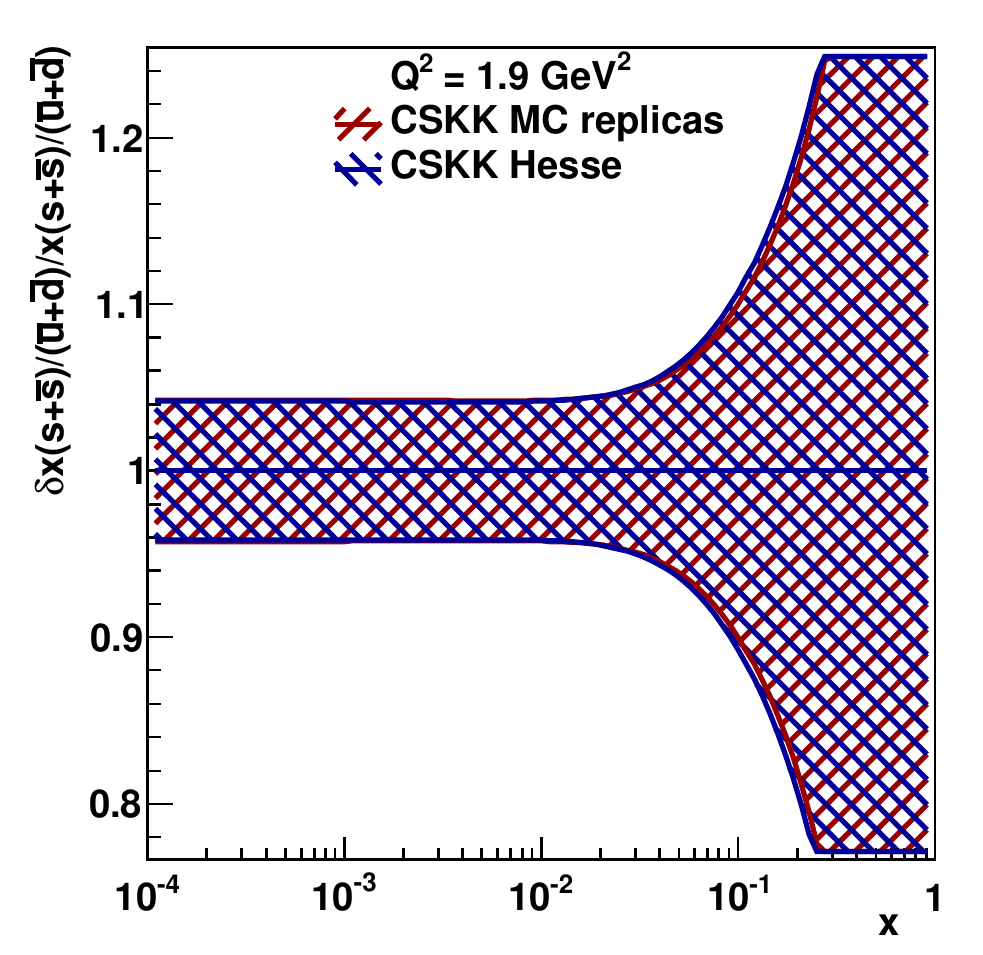}}
  \vspace{.5cm}
  \caption {\large PDF experimental uncertainties for the $u$ and $d$ valence, total Sea $\Sigma$ and 
    the ratio $(s +\bar{s})/(\bar{u} +\bar{d}) $ for the CSKK fit where the
    uncertainties are calculated using Hesse and MC replicas method, respectively.
    Full details are given in the text.
  }
  \label{fig:fig9}
\end{figure}

\clearpage

\begin{figure}[tbp]
`  \vspace{-0.3cm}
  \centerline{
    \includegraphics[width=0.50\textwidth]{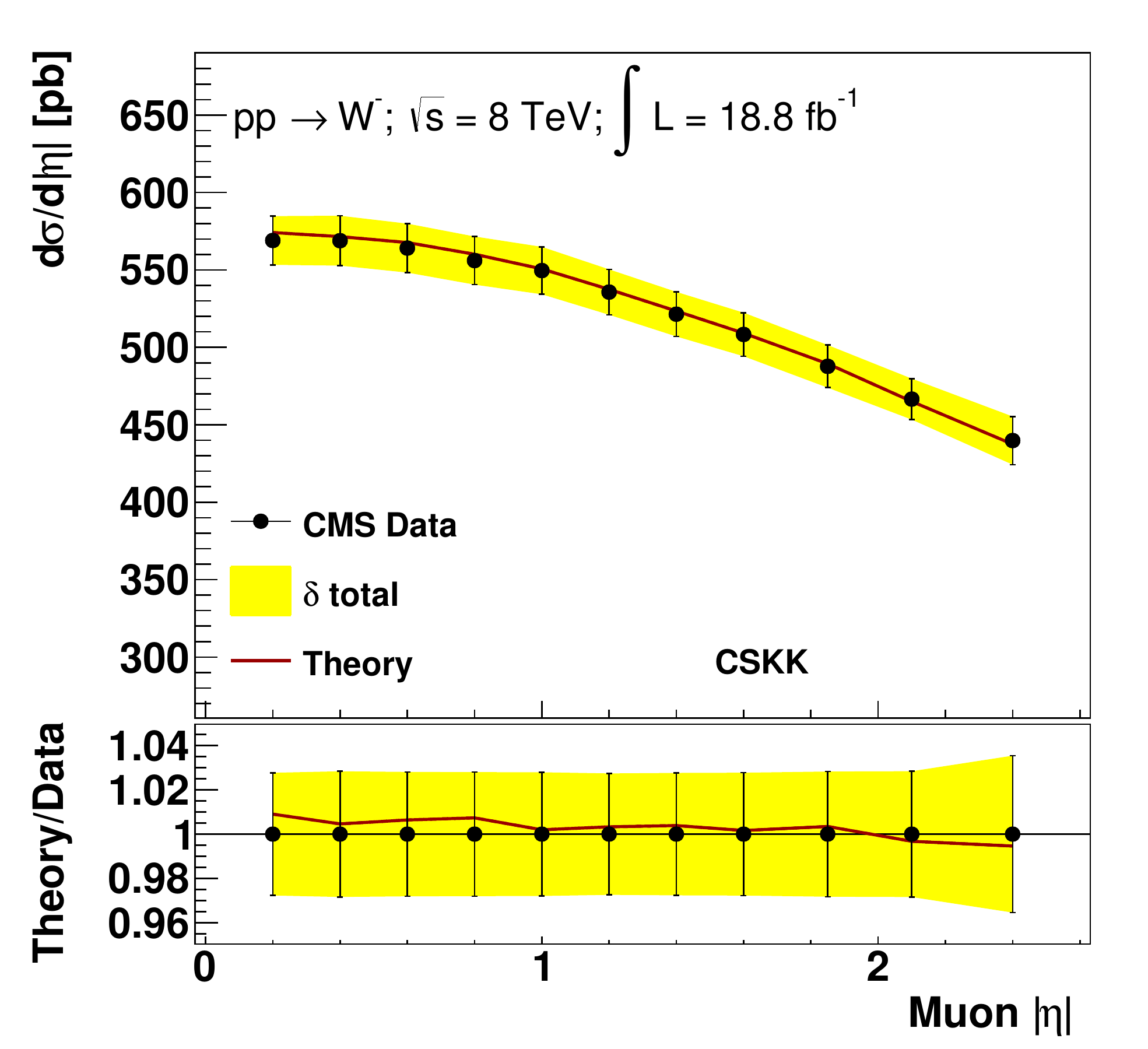}
    \includegraphics[width=0.50\textwidth]{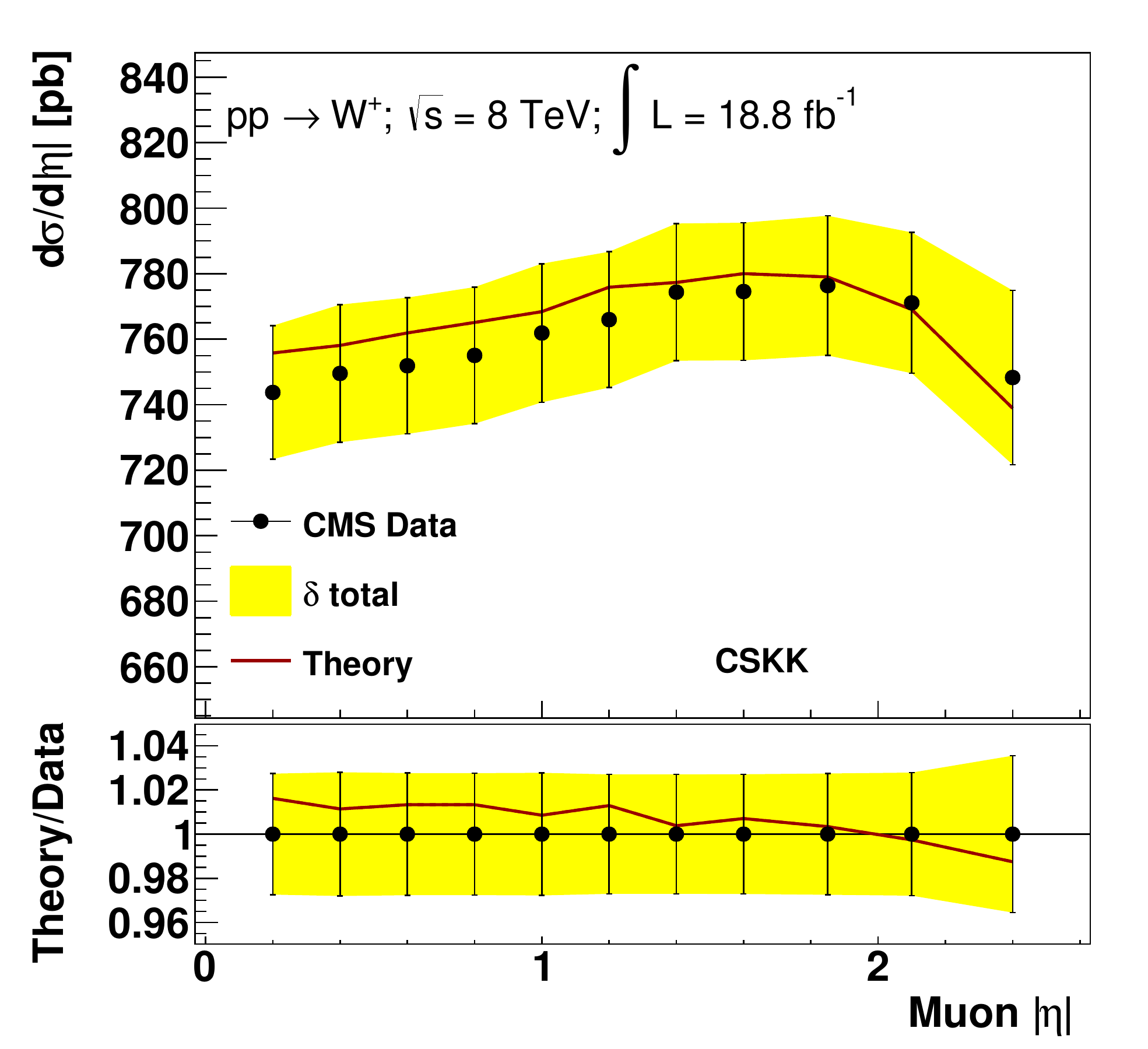}}
  \vspace{0.5cm}
  \caption {\large The PDF fit (Theory, CSKK) to ATLAS, CMS and HERA data compared to the CMS $W^+$ data (left)
    and $W^-$ data at 8 TeV.
  }
  \label{fig:fig10}
\end{figure}
\begin{figure}[tbp]
  \vspace{-0.3cm}
  \centerline{
    \includegraphics[width=0.50\textwidth]{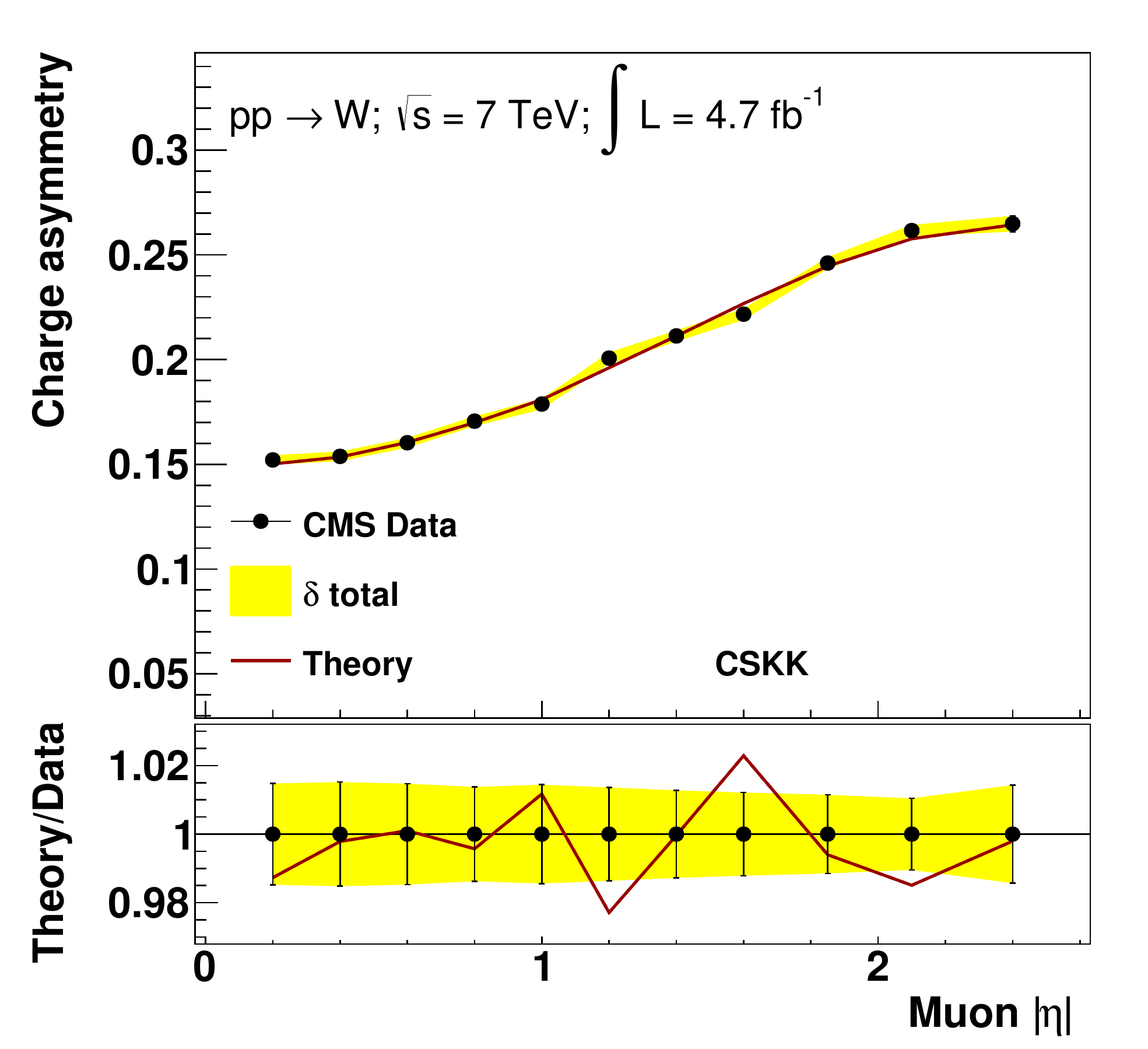}}
  \vspace{0.5cm}
  \caption {\large The PDF fit (Theory, CSKK) to ATLAS, CMS and HERA data compared to CMS $W$-asymmetry data at 7 TeV. 
  }
  \label{fig:fig11}
\end{figure}

\clearpage

\begin{figure}[tbp]
  \vspace{-0.3cm}
  \centerline{
    \includegraphics[width=0.50\textwidth]{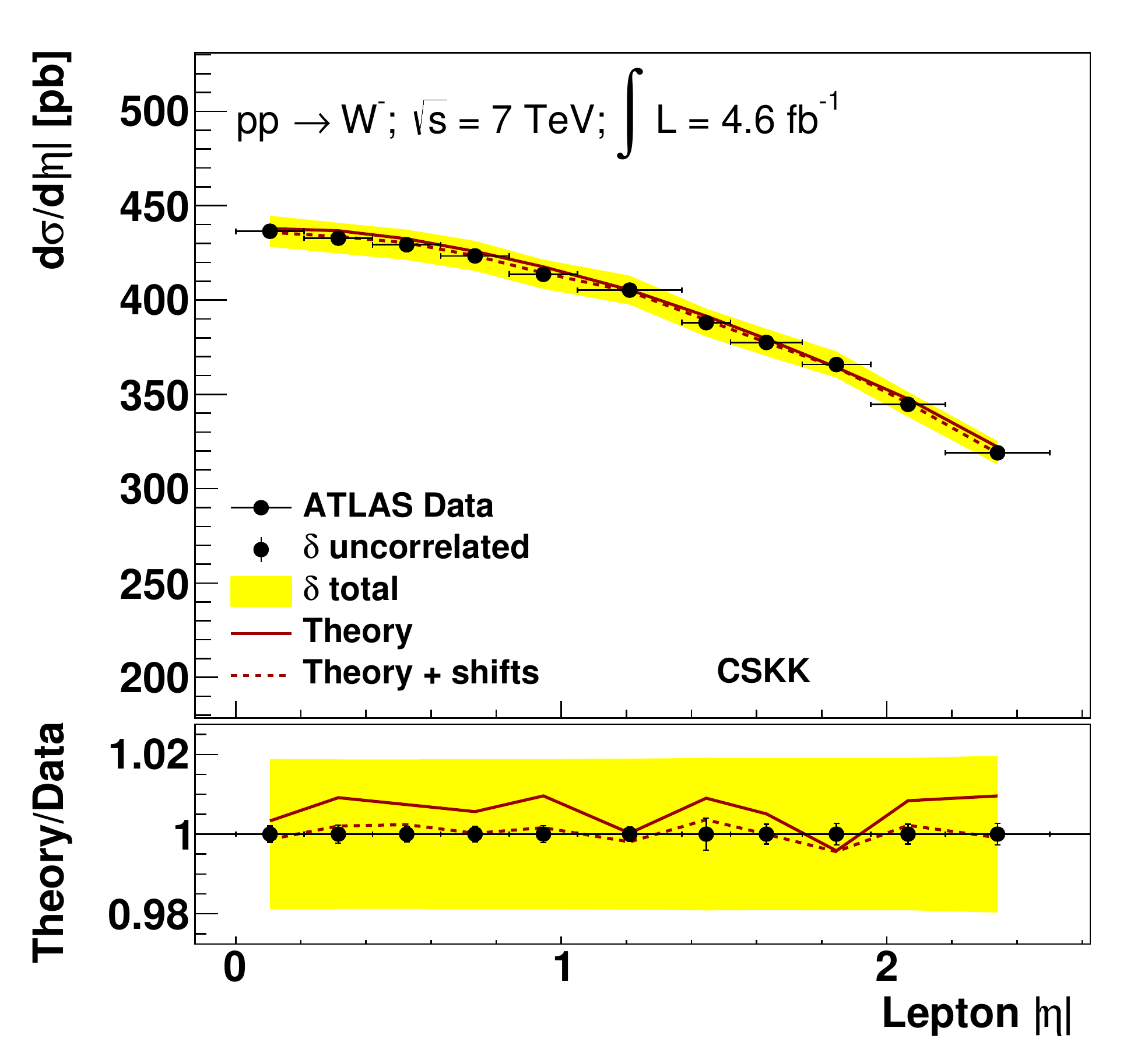}
    \includegraphics[width=0.50\textwidth]{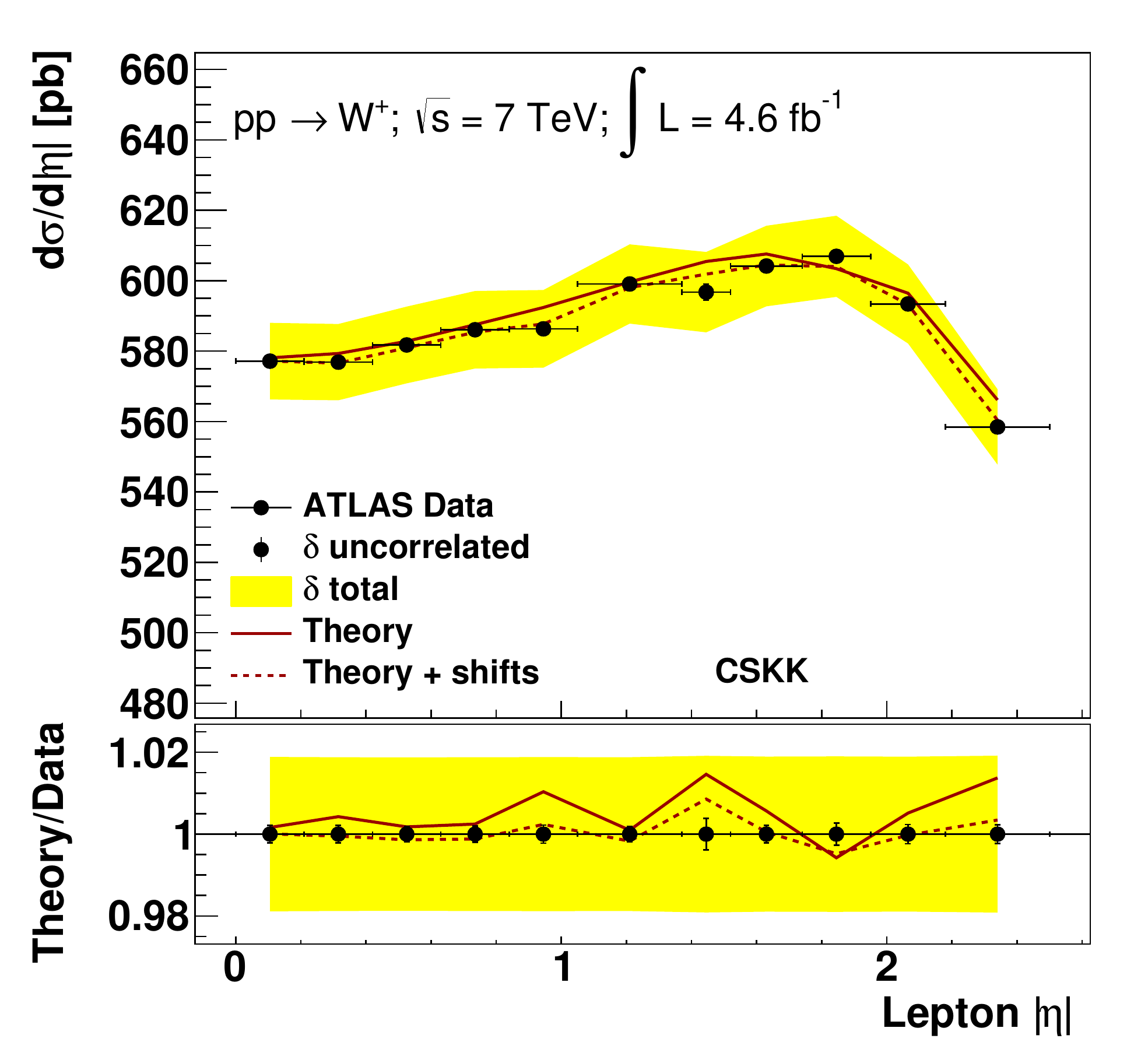}}
  \vspace{0.5cm}
  \caption {\large The PDF fit (Theory, CSKK) to ATLAS, CMS and HERA data compared to 
ATLAS  $W^+$ data (left) and $W^-$ data (right) at 7 TeV. Since the correlated systematic 
uncertainties for ATLAS are allowed to vary, by shifts determined by nuisance parameters 
in the fit, the theoretical prediction is also shown after the shifts (Theory + shifts). 
  }
  \label{fig:fig12}
\end{figure}

\clearpage

\begin{figure}[tbp]
  \vspace{-0.3cm}
  \centerline{
    \includegraphics[width=0.50\textwidth]{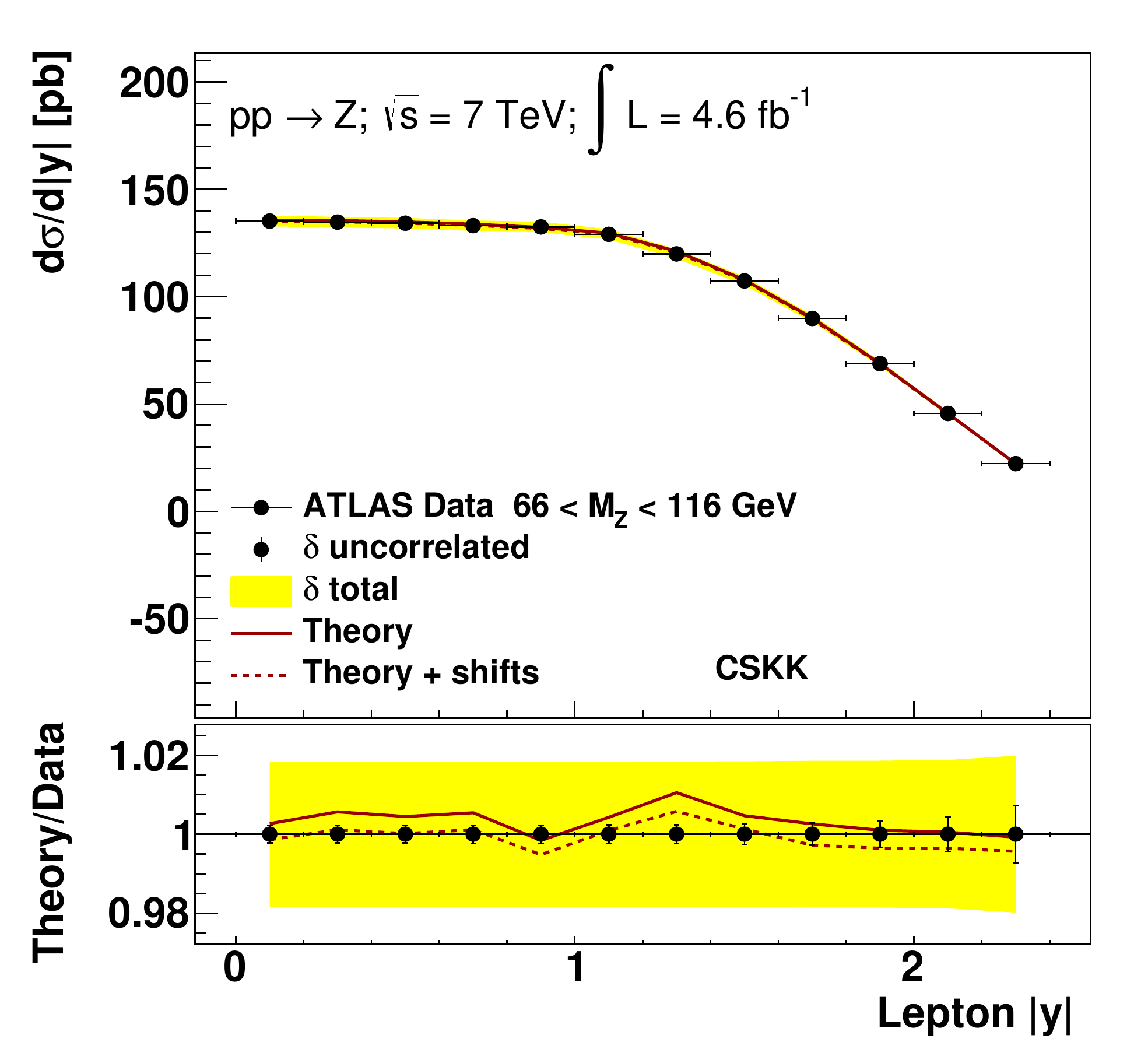}
    \includegraphics[width=0.50\textwidth]{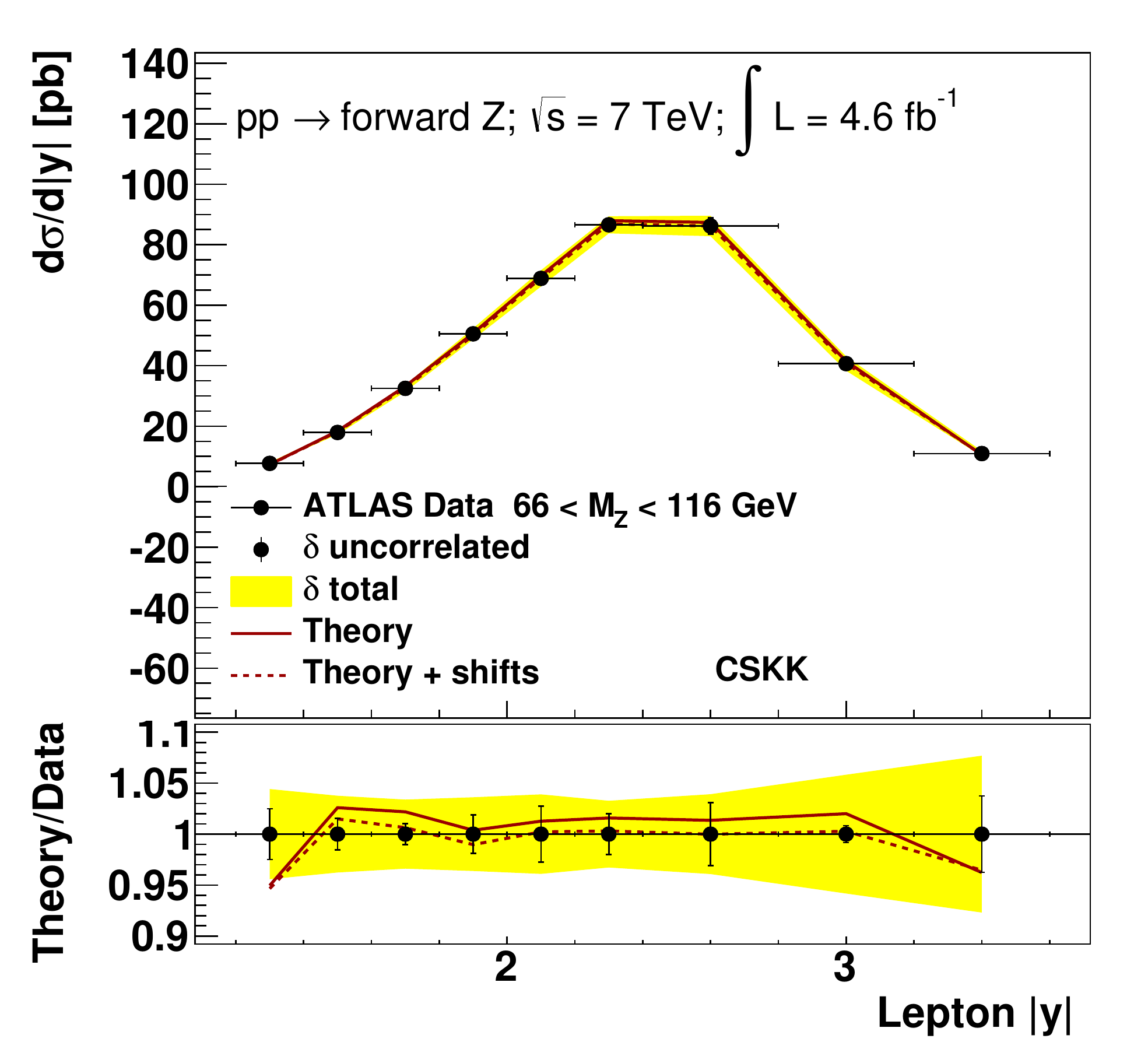}}
  \vspace{0.5cm}
  \caption {\large The PDF fit (Theory, CSKK) to ATLAS, CMS and HERA data compared to ATLAS $Z$ data at central rapidity (left) and forward rapidity (right) at 7 TeV. Since the correlated systematic 
uncertainties for ATLAS are allowed to vary, by shifts determined by nuisance parameters 
in the fit, the theoretical prediction is also shown after the shifts (Theory + shifts).
  }
  \label{fig:fig13}
\end{figure}
\begin{figure}[tbp]
  \vspace{-0.3cm}
  \centerline{
    \includegraphics[width=0.50\textwidth]{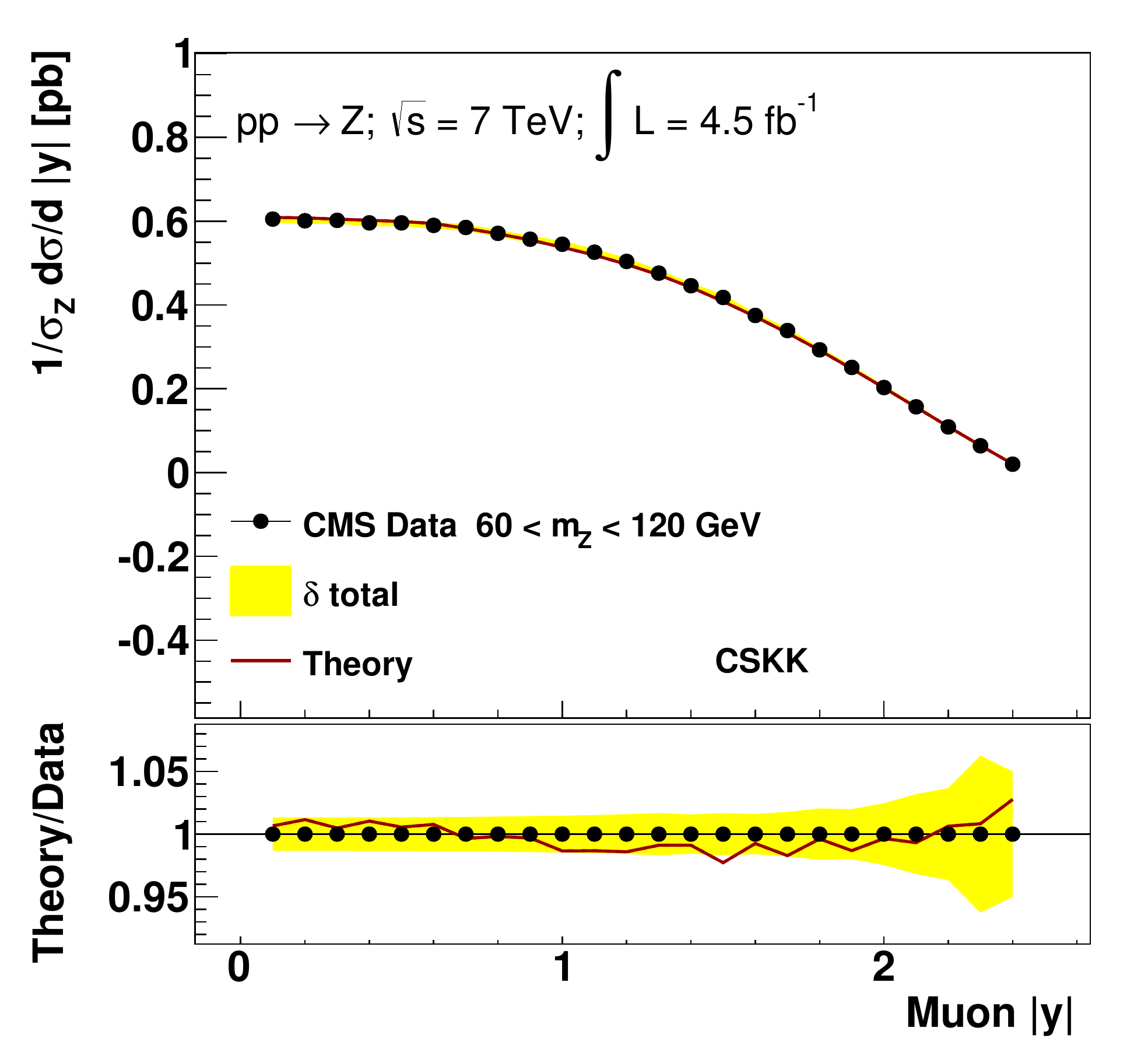}}
  \vspace{0.5cm}
  \caption {\large The PDF fit (Theory, CSKK) to ATLAS, CMS and HERA data compared to CMS $Z$ data at 7 TeV. 
  }
  \label{fig:fig14}
\end{figure}

\clearpage
\begin{figure}[tbp]
  \vspace{-5.5cm}
  \centerline{
    \includegraphics[width=1.1\textwidth]{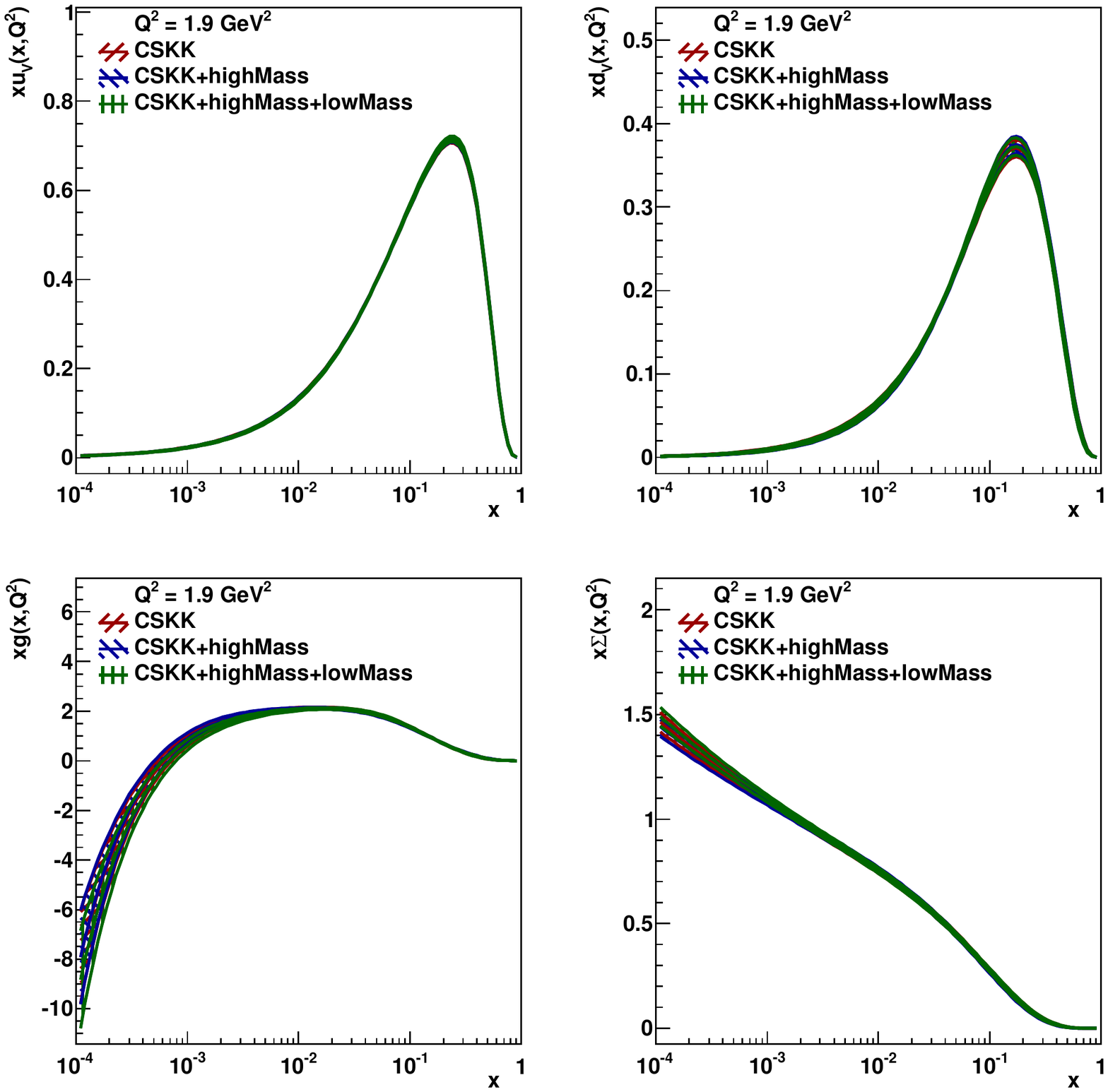}}
  \vspace{-2.5cm}
  \caption {\large The PDFs $u_v$, $d_v$, gluon and total sea for the nominal CSKK fit to HERA data 
and ATLAS and CMS $W$ and $Z$ mass-peak data, compared to fits which include the 7 TeV high and 
low mass off-peak $Z$ data. The bands represent experimental uncertainties. Full details are given in the text.
  }
  \label{fig:fig15}
\end{figure}

\clearpage

\begin{figure}[tbp]
  \vspace{-5.5cm}
  \centerline{
    \includegraphics[width=1.1\textwidth]{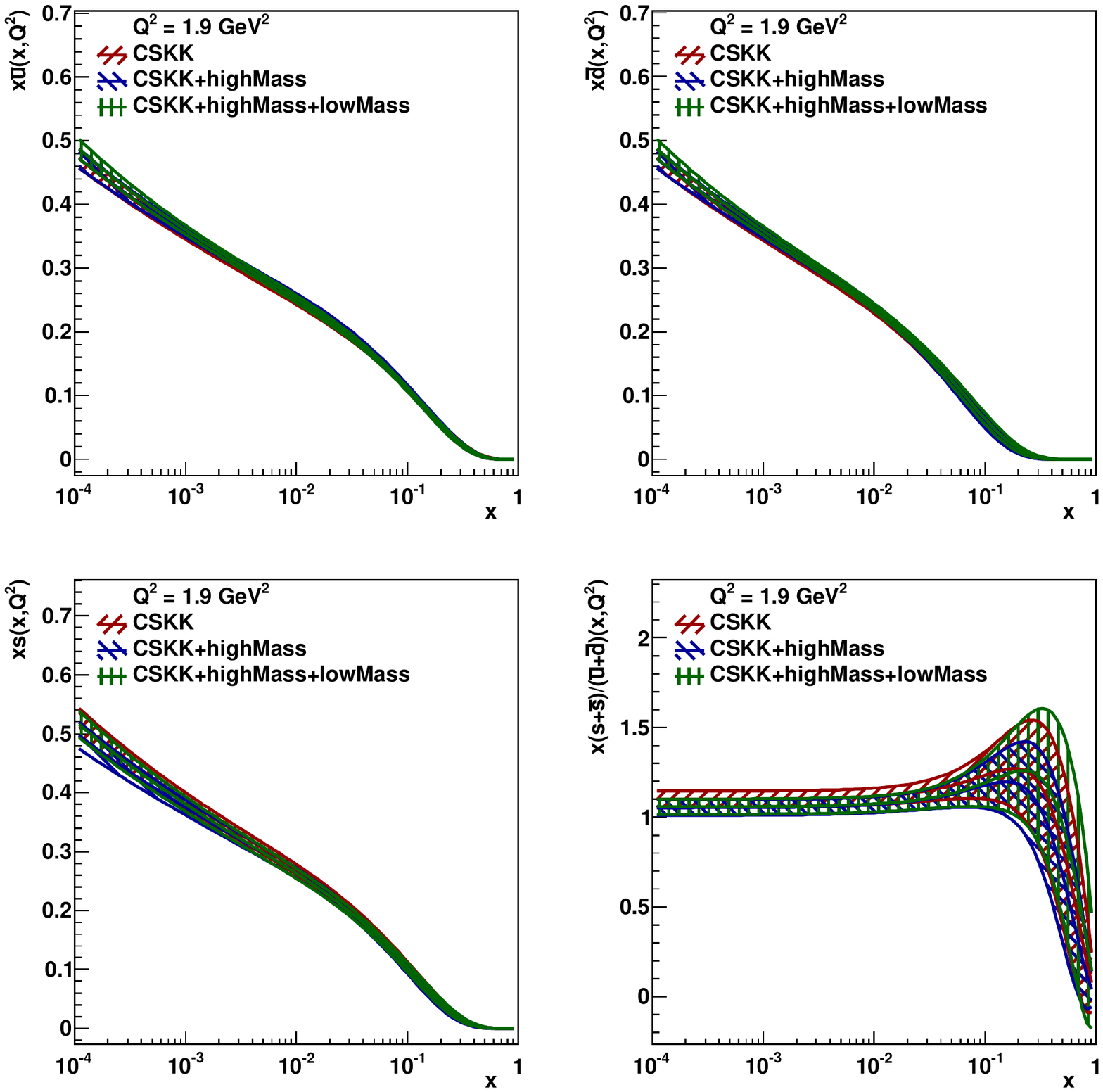}}
  \vspace{-2.5cm}
  \caption {\large  The PDFs $\bar{u}$, $\bar{d}$, $s$ and 
the ratio $(s +\bar{s})/(\bar{u} +\bar{d}) $ for the nominal CSKK fit to HERA data 
and ATLAS and CMS $W$ and $Z$ mass-peak data, compared to fits which include the 7 TeV high and 
low mass off-peak $Z$ data. The bands represent experimental uncertainties. 
Full details are given in the text.
  }
  \label{fig:fig16}
\end{figure}

\clearpage

\begin{figure}[tbp]
  \vspace{-0.3cm}
  \centerline{
    \includegraphics[width=0.5\textwidth]{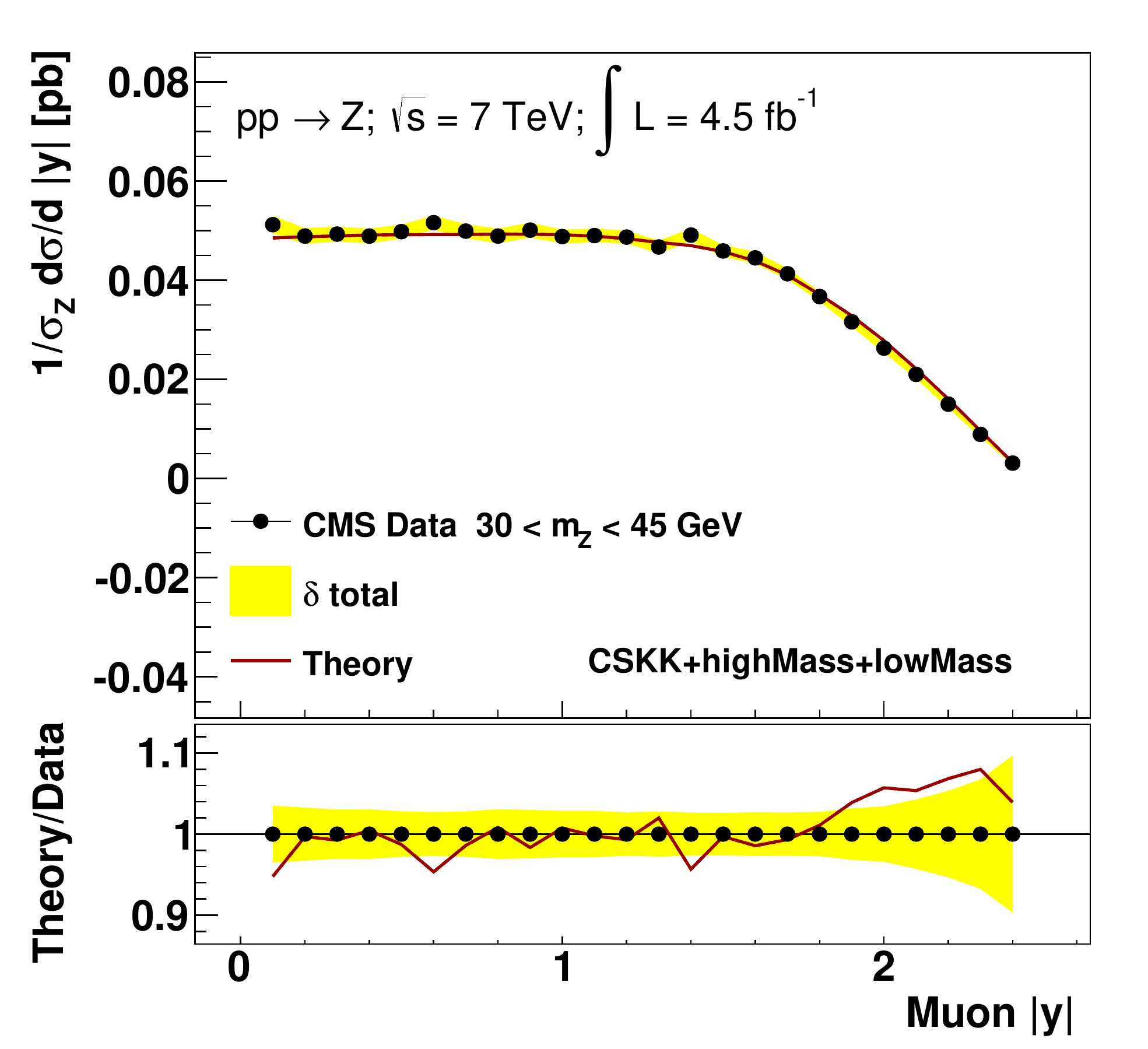}
    \includegraphics[width=0.5\textwidth]{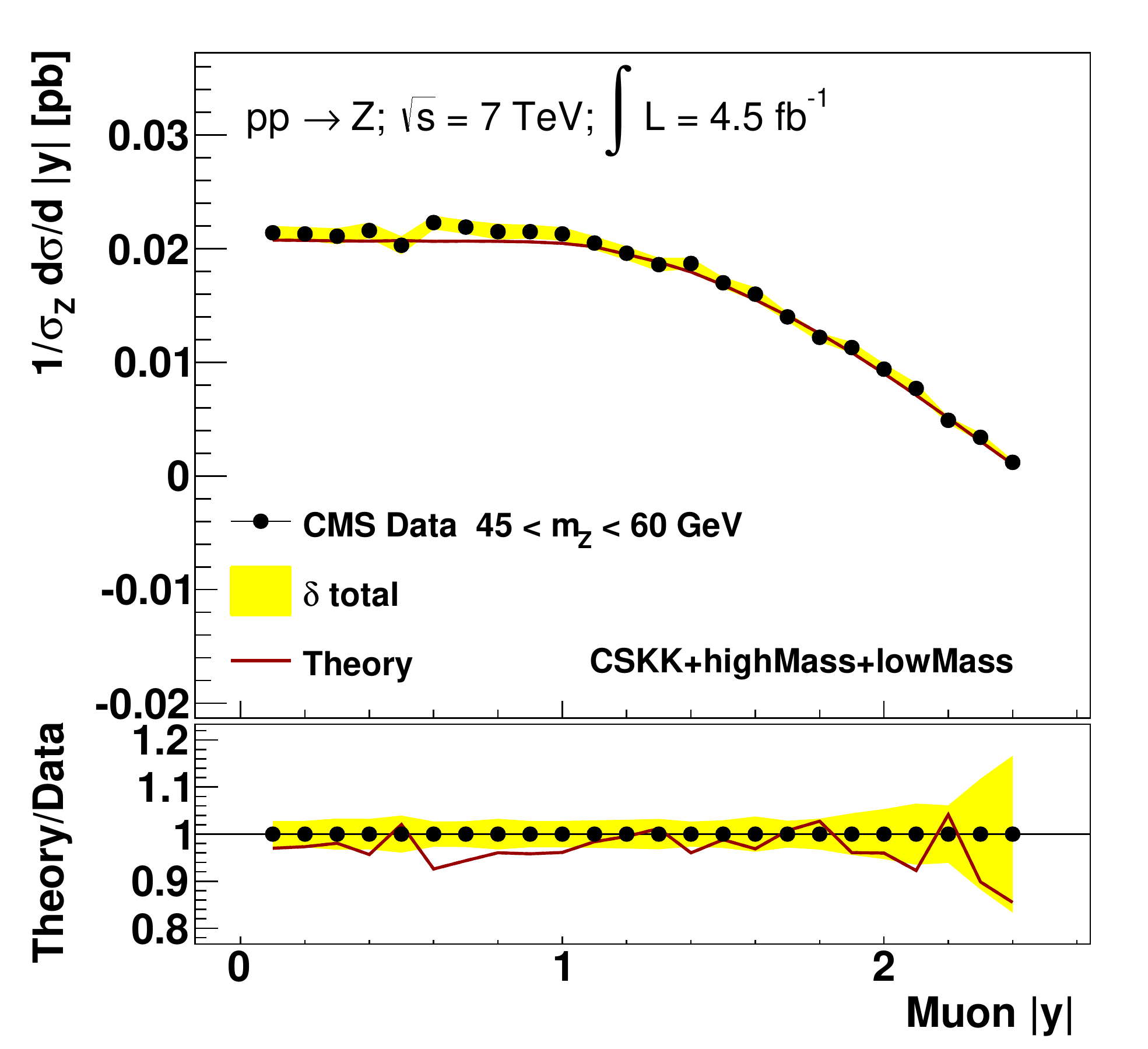}}
\end{figure}
\begin{figure}[tbp]
  \vspace{-5.3cm}
  \centerline{
    \includegraphics[width=0.5\textwidth]{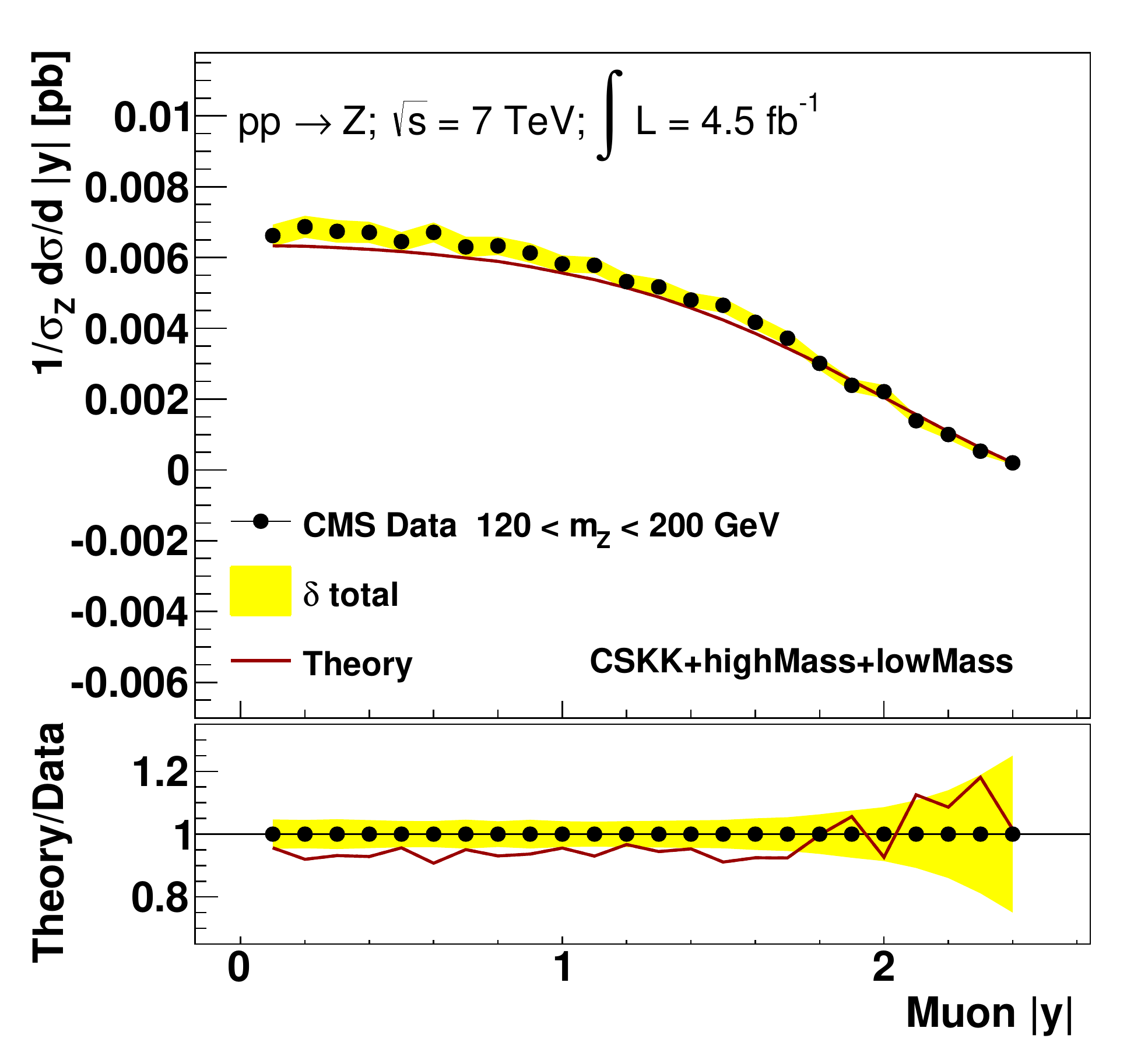}
    \includegraphics[width=0.5\textwidth]{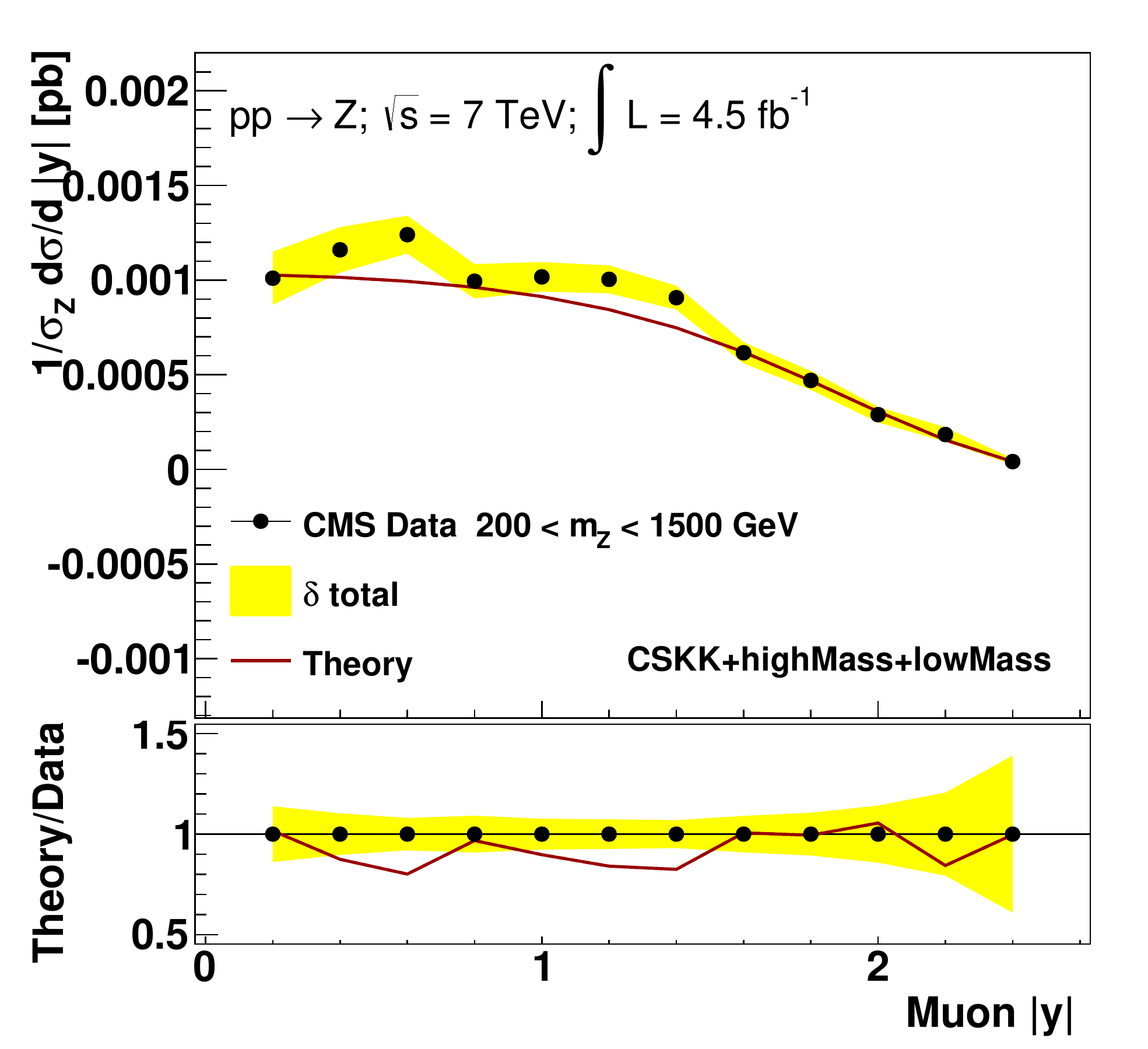}}
  \vspace{0.5cm}
  \caption {\large The PDF fit (Theory, CSKK+highMass+lowMass) to HERA and ATLAS and CMS data for all $Z$ mass regions compared
    to CMS off-peak $Z$ data at 7 TeV
  }
  \label{fig:fig17}
\end{figure}

\clearpage
\begin{figure}[tbp]
  \vspace{-0.3cm}
  \centerline{
    \includegraphics[width=0.5\textwidth]{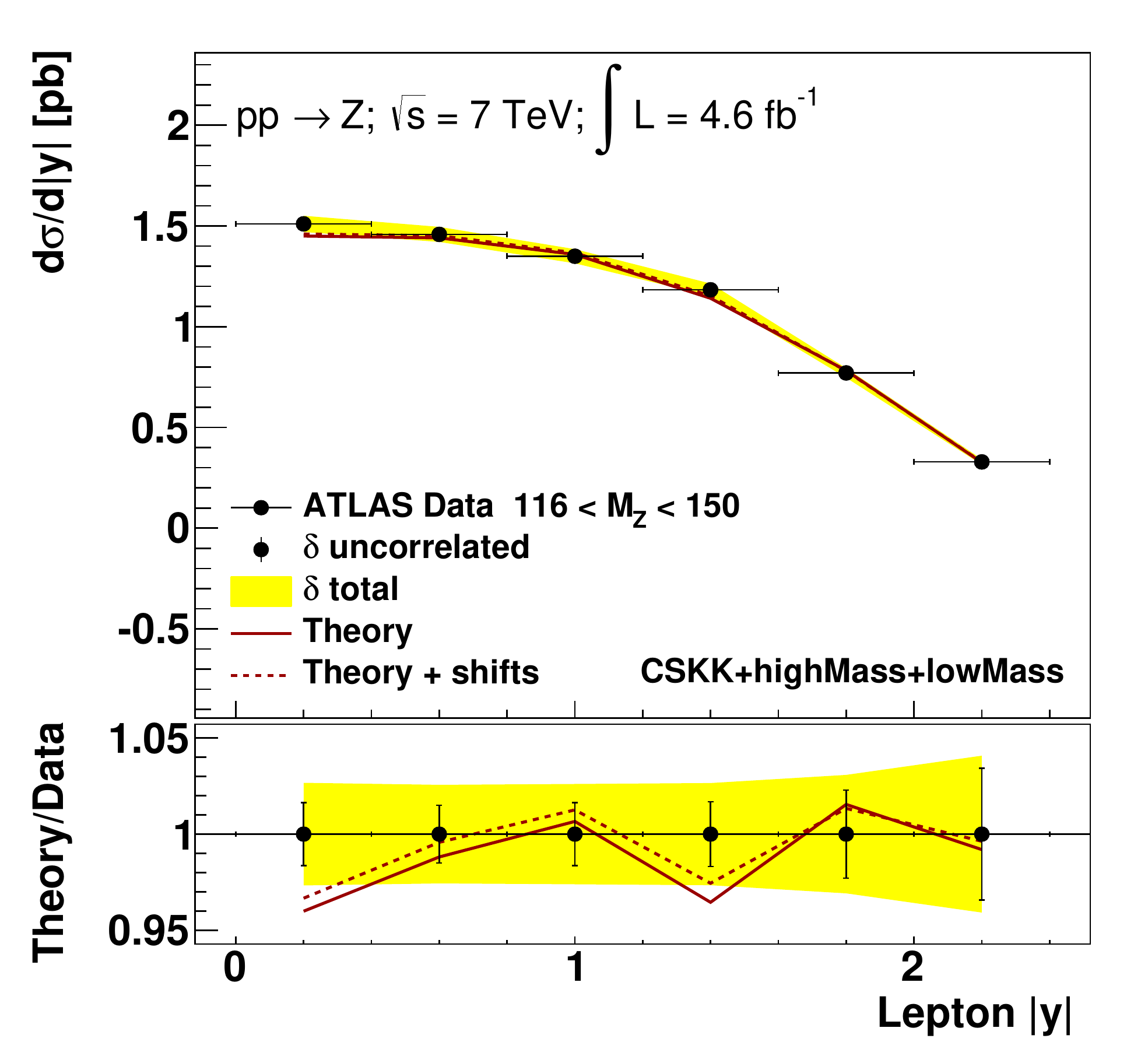}
    \includegraphics[width=0.5\textwidth]{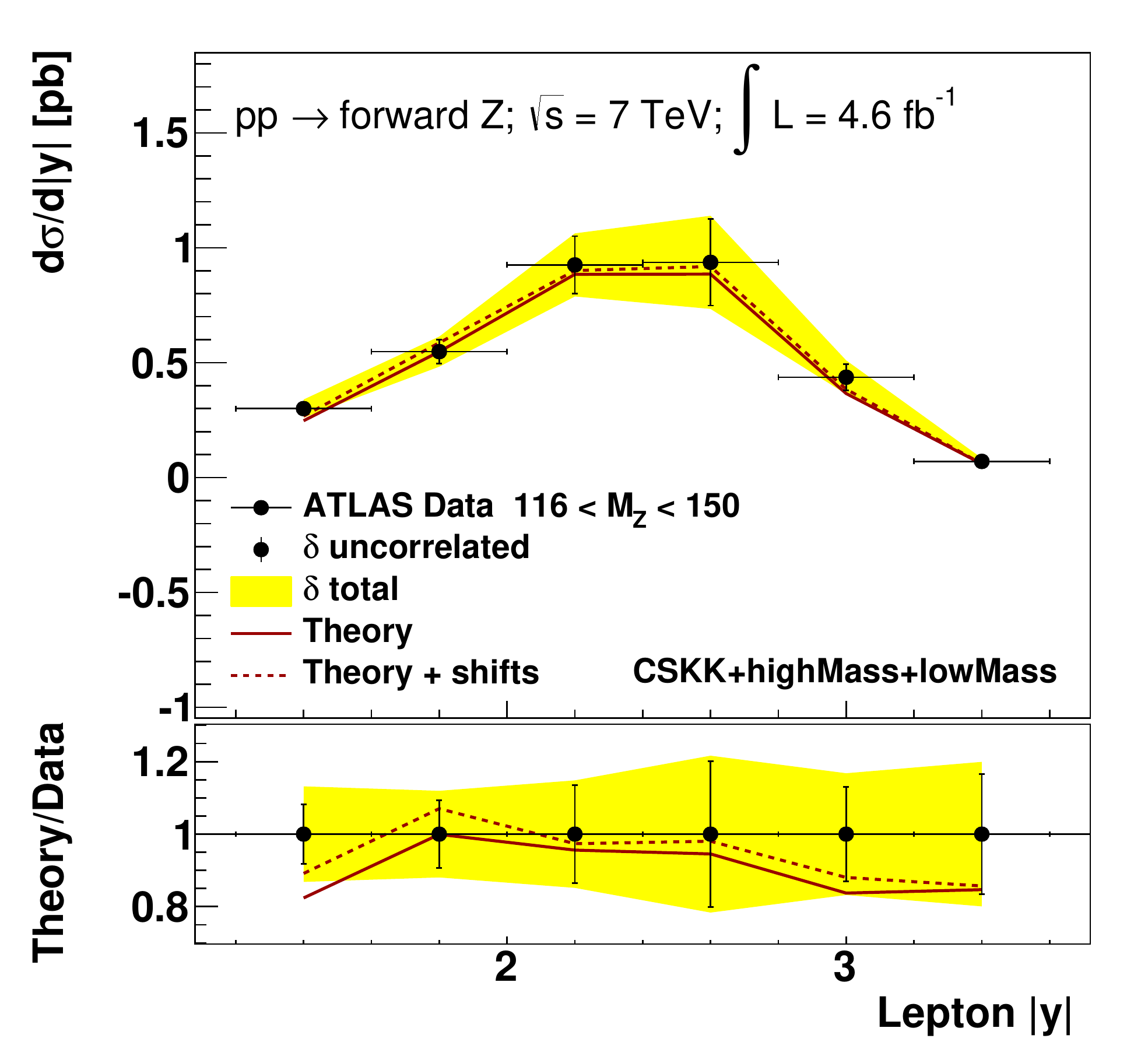}}
\end{figure}
\begin{figure}[tbp]
  \vspace{-5.3cm}
  \centerline{
    \includegraphics[width=.5\textwidth]{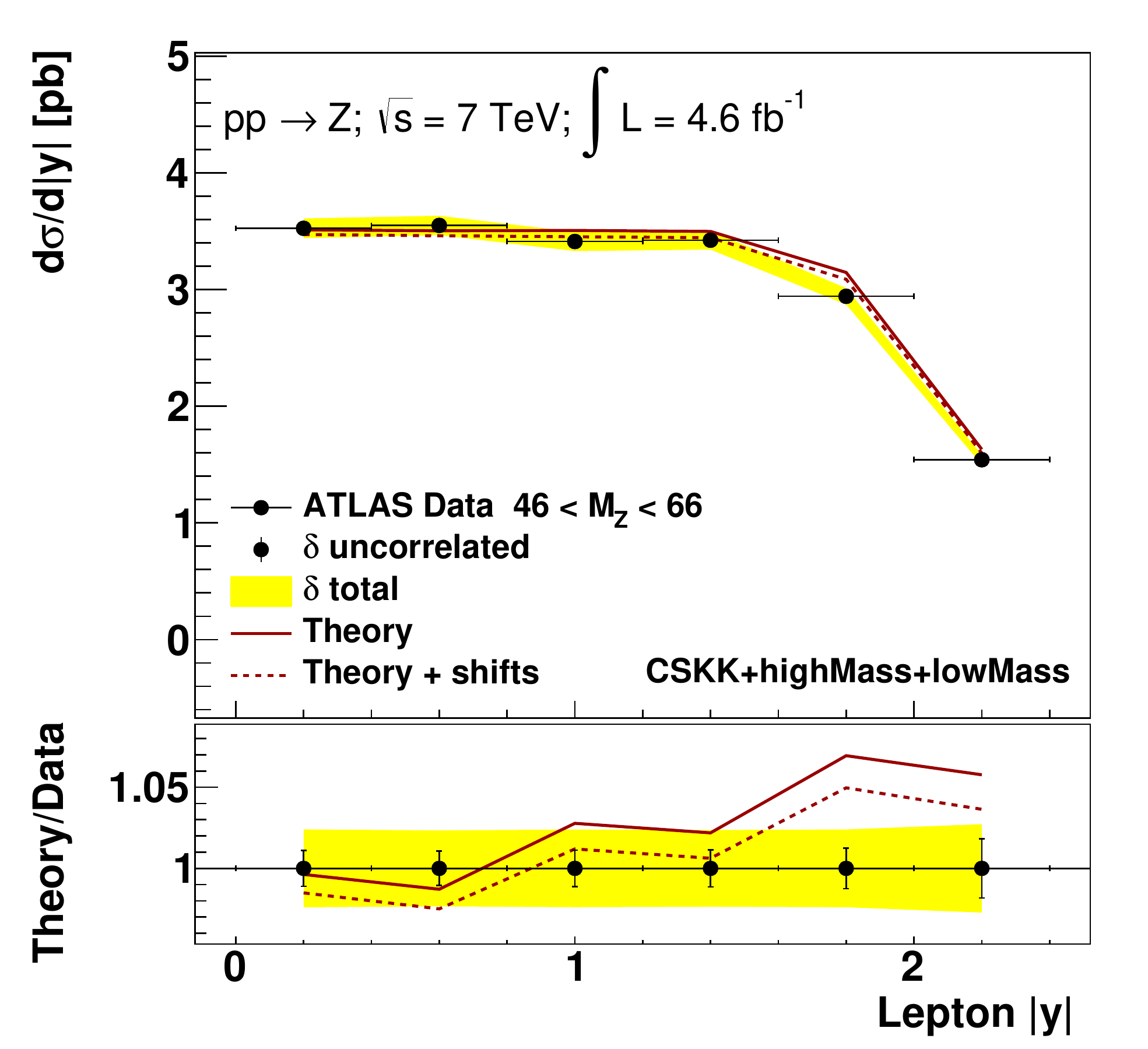}}
  \vspace{0.5cm}
  \caption {\large
    The PDF fit (Theory, CSKK+highMass+lowMass) to HERA and ATLAS and CMS data for all $Z$ mass regions compared to ATLAS off-peak $Z$ data at 7 TeV. Since the correlated systematic 
uncertainties for ATLAS are allowed to vary, by shifts determined by nuisance parameters 
in the fit, the theoretical prediction is also shown after the shifts (Theory + shifts).
}
  \label{fig:fig18}
\end{figure}

\clearpage
\begin{figure}[tbp]
  \vspace{-5.5cm}
  \centerline{
    \includegraphics[width=1.07\textwidth]{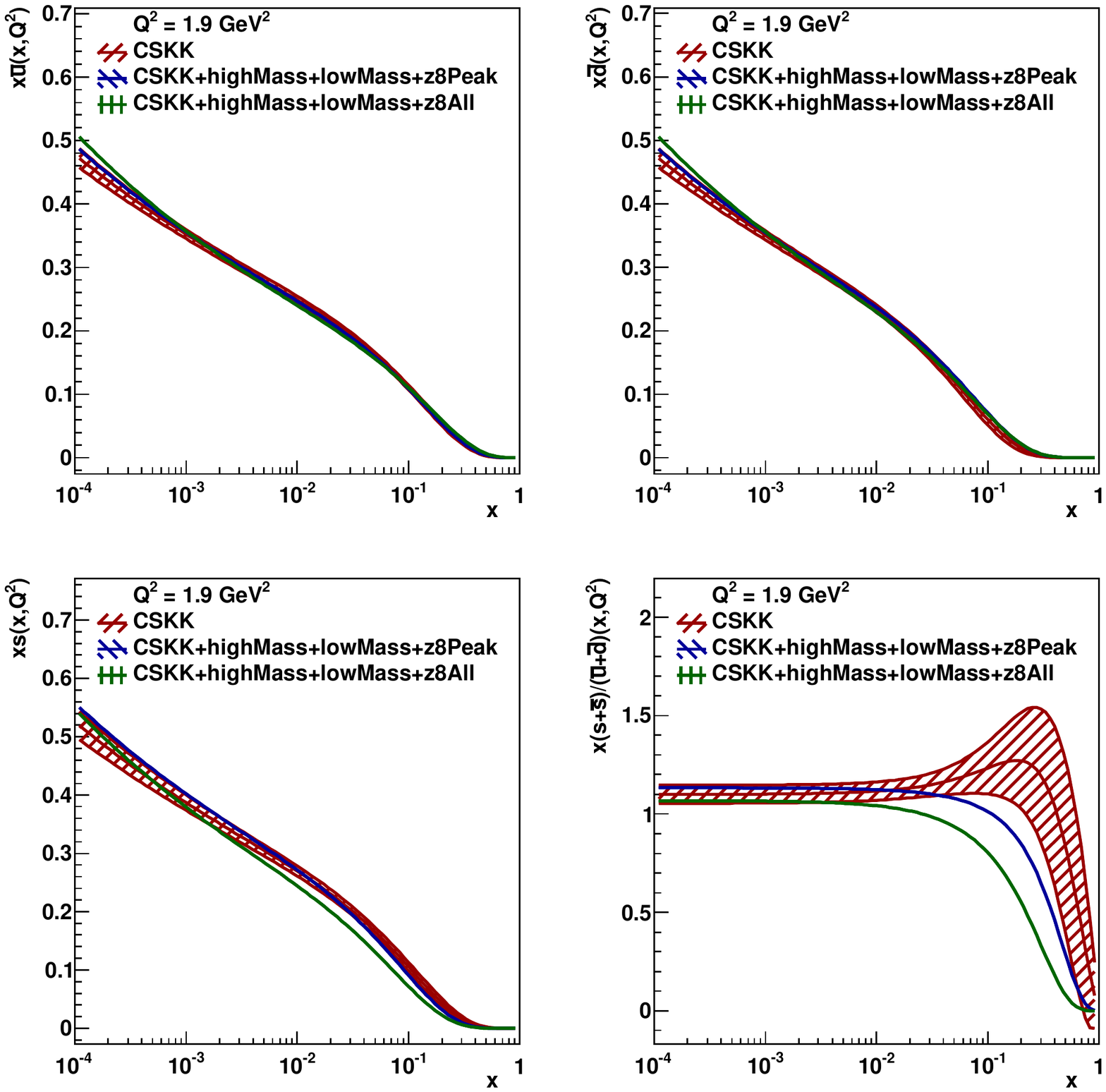}}
  \vspace{-2.5cm}
  \caption {\large  The central values of the PDFs $\bar{u}$, $\bar{d}$, $s$ and 
the ratio $(s +\bar{s})/(\bar{u} +\bar{d}) $ for fits to HERA data
and ATLAS and CMS $W$ and $Z$ data including the high and low di-lepton invariant mass 
regions at 7 TeV, as well as the CMS 8 TeV $Z$ data. These are compared to the nominal CSKK fit 
and its experimental uncertainties. Full details are given in the text. 
}
  \label{fig:fig19}
\end{figure}

\clearpage
\begin{figure}[tbp]
  \vspace{-5.5cm}
  \centerline{
    \includegraphics[width=1.1\textwidth]{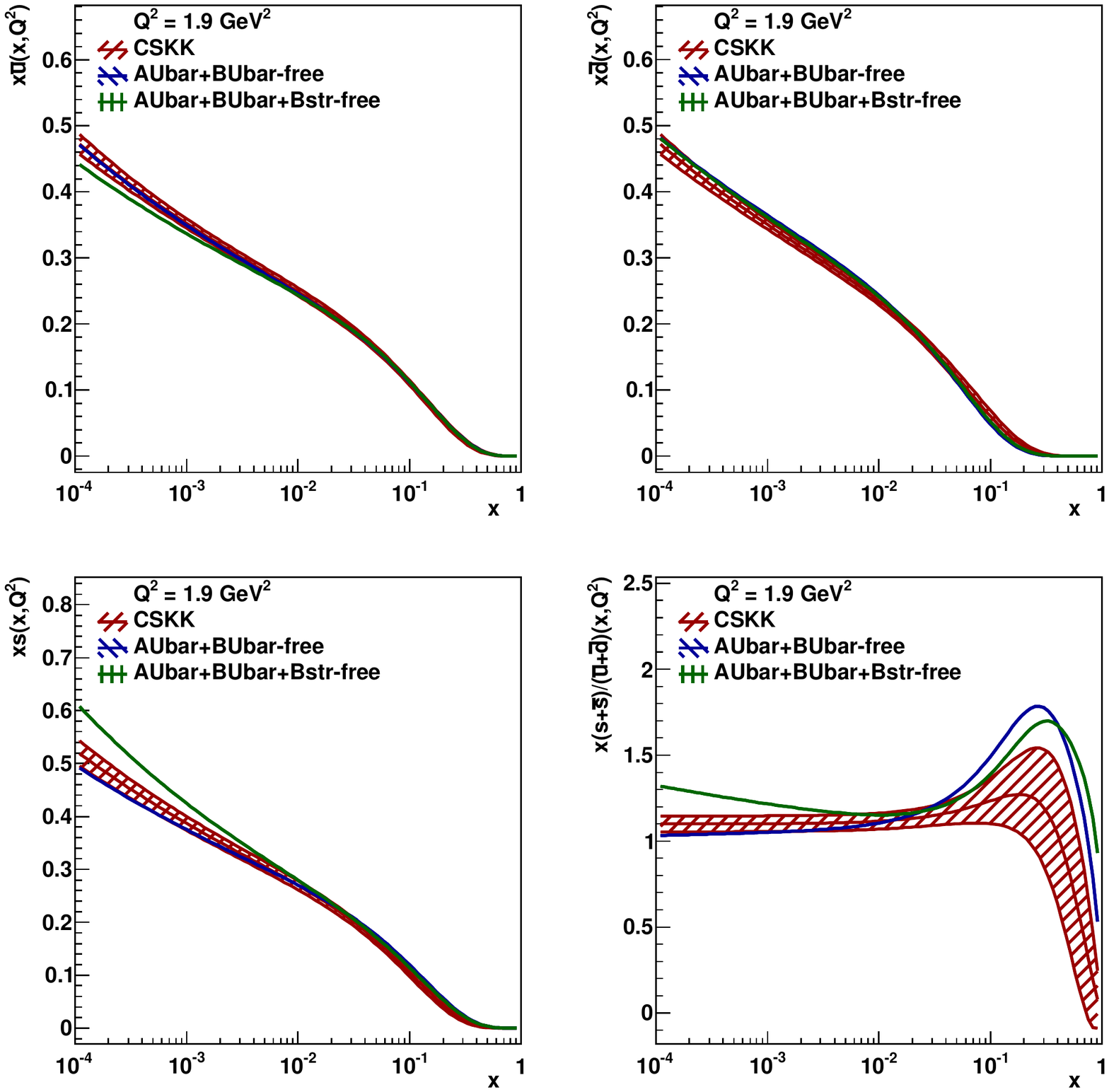}}
  \vspace{-2.5cm}
  \caption {\large  The PDFs $\bar{u}$, $\bar{d}$, $s$ and 
the ratio $(s +\bar{s})/(\bar{u} +\bar{d}) $ for the nominal CSKK fit to HERA data 
and ATLAS and CMS $W$ and $Z$ mass-peak data, compared to fits with additional free 
parameters for the $\bar{u}, \bar{d}$ and $s$ sea. 
The bands represent experimental uncertainties for the CSKK fit.
Full details are given in the text.
  }
  \label{fig:fig20}
\end{figure}

\clearpage
\begin{figure}[tbp]
  \vspace{-.5cm}
  \centerline{
        \includegraphics[width=.65\textwidth]{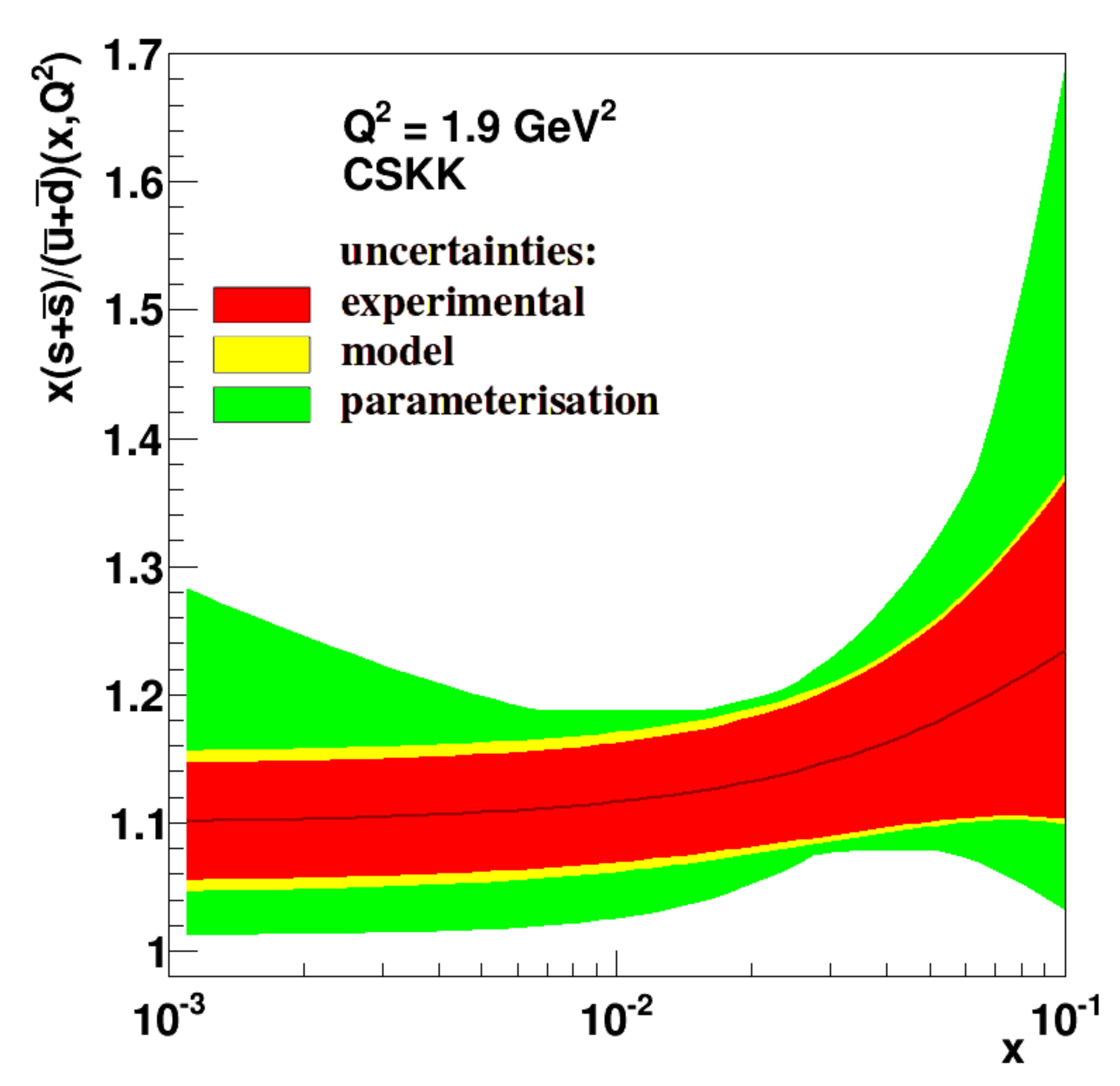}}    
\end{figure}
\begin{figure}[tbp]
  \vspace{-3.cm}
  \centerline{
    \includegraphics[width=.65\textwidth]{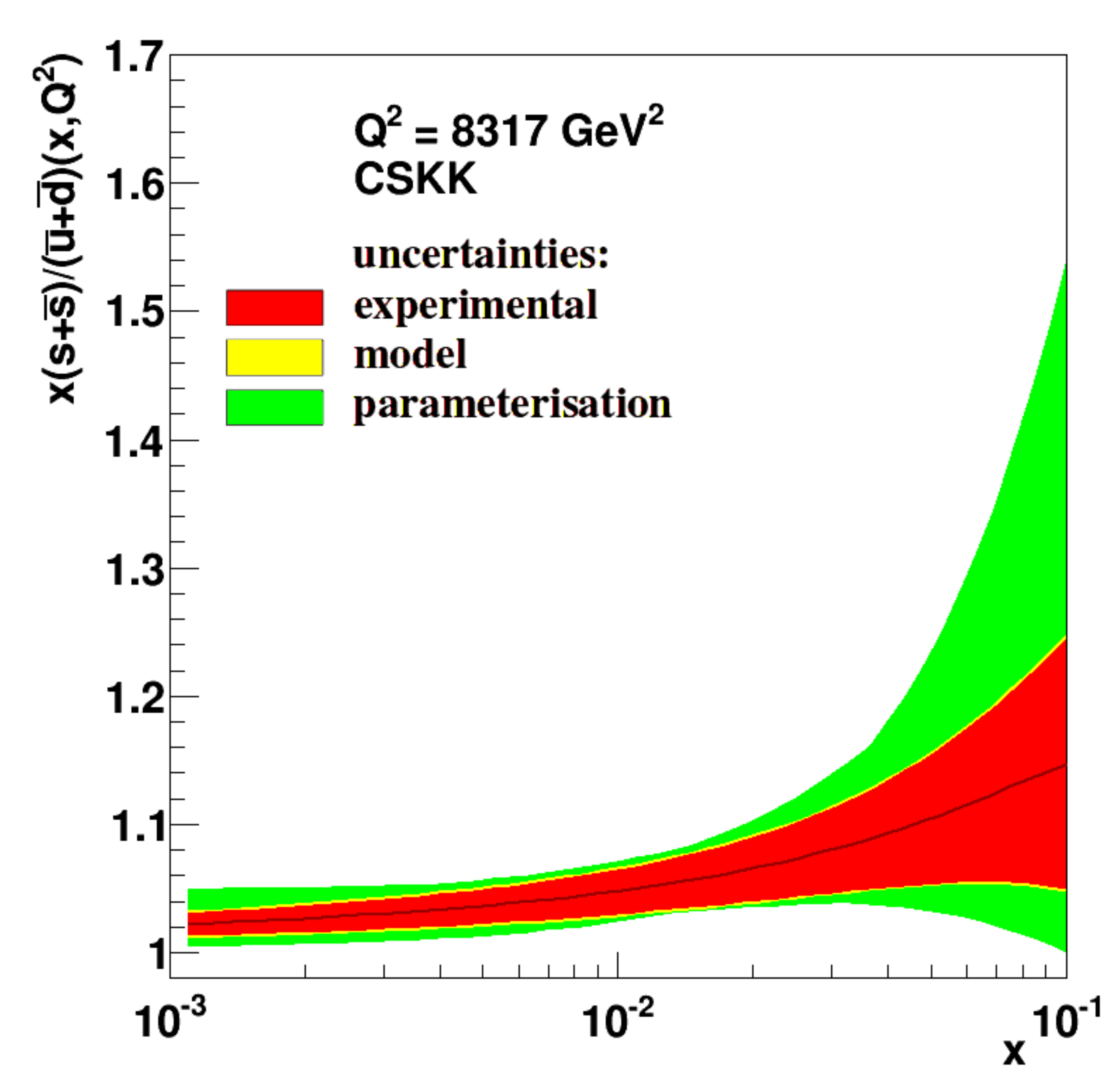}}
  \caption {\large The ratio $(s +\bar{s})/(\bar{u} +\bar{d}) $ for fits to HERA data
    and ATLAS and CMS $W$ and $Z$ data with experimental, model and parameterisation
    uncertainties at the starting scale $Q^2_0$ (top) and at $Q^2=M^2_Z$ (bottom).
    Full details are given in the text. 
}
  \label{fig:fig21}
\end{figure}


\vspace{2.5cm}
\begin{thebibliography}{0}
\bibitem{1506.06042}
  H.  Abramowicz et al., [ZEUS  and  H1  Collaboration],  Eur.  Phys.  J.  C75,  580  (2015), [arXiv:1506.06042].
\bibitem{JimenezDelgado:2008hf}
  P. Jimenez-Delgado and E. Reya, Phys. Rev. D79 (2009) 074023, [arXiv:0810.4274].
\bibitem{HERA:2009wt}
  F.D. Aaron et al., [H1 and ZEUS Collaborations],  JHEP01 (2010) 109, [arxiv:0911.0884].
\bibitem{Adloff:2000qk}
  C. Adloff et al., [H1 Collaboration], Eur. Phys. J. C 21 (2001) 33-61, [hep-ex/0012053].
\bibitem{PhysRevD.74.012008}
  M. Tzanov et al., [NuTeV Collaboration], Phys.Rev. D74 (2006) 012008, [hep-ex/0509010].
\bibitem{Goncharov:2001qe}
  M. Goncharov et al., Phys. Rev. D 64 (2001) 112006, [hep-ex/0102049].
\bibitem{Samoylov:2013}
  O. Samoylov et al., Nucl. Phys. B 876 (2013) 339, [arXiv:1308.4750].
\bibitem{1612.03016}
  ATLAS Collaboration, Eur. Phys. J. C 77 367 (2017), [arXiv:1612.03016]. \\
  data files and K-factors from  https://www.hepdata.net/record/76541 \\
  APPLGRID calculations from \\
  https://atlas.web.cern.ch/Atlas/GROUPS/PHYSICS/PAPERS/STDM-2012-20
\bibitem{atlaswcharm}
  ATLAS Collaboration, JHEP 05 (2014) 068, [arXiv:1402.6263].
\bibitem{cmswcharm}
  CMS Collaboration, JHEP 02 (2014) 013, [arXiv:1310.1138].
\bibitem{1312.6283}
  CMS Collaboration, Phys. Rev. D 90 (2014) 032004, [arXiv:1312.6283]. \\
  data files, K-factors and APPLGRID calculations from
  http://www.hepforge.org/archive/xfitter/1312.6283.tar.gz
\bibitem{1603.01803}
  CMS Collaboration, Eur. Phys. J. C 76 (2016) 469, [arXiv:1603.01803]. \\
  data files, K-factors and APPLGRID calculations from
  http://www.hepforge.org/archive/xfitter/1603.01803.tar.gz
\bibitem{1310.7291}
  CMS Collaboration, JHEP 12 (2013) 030, [arXiv:1310.7291].  \\
  data files from https://hepdata.net/record/ins1262319
\bibitem{1412.1115}
  CMS Collaboration, Eur. Phys. J. C 75 (2015) 147, [arXiv:1412.1115]. \\
  data files from https://hepdata.net/record/ins1332509
\bibitem{nnpdf31}
 R. Ball et al., [NNPDF Collaboration], submitted to EPJC, [arXiv:1706.00428].
\bibitem{HERAFitter}
  S. Alekhin et al. (2014), [arXiv:1410.4412].
\bibitem{Aaron:2009kv}
  F. D. Aaron et al., [H1 Collaboration], Eur. Phys. J. C64:561-587 (2009), [arXiv:0904.3513].
\bibitem{Botje:2010ay}
  M. Botje, Comp. Phys. Comm. 182, 490 (2011), [arXiv:1005.1481].
\bibitem{Thorne:1997ga}
 R. S. Thorne and R. G. Roberts, Phys. Rev. D57, 6871 (1998), [hep-ph/9709442].\\
 R. S. Thorne, Phys. Rev. D73, 054019 (2006), [hep-ph/0601245].  
\bibitem{Thorne:2006qt}
R. S. Thorne, Phys. Rev. D86, 074017 (2012), [arXiv:1201.6180].
\bibitem{minuit}
F. James and M. Roos, Comput.Phys.Commun. 10 (1975) 343-367.
\bibitem{zeusfitter}
  S.Chekanov et al., [ZEUS Collaborations], Phys. Rev. D67, 012007 (2003) [hep-ex/0208023]
\bibitem{xFittermanual}
  The documentation and the package can be found at: URL https://www.xfitter.org/xFitter/ .
\bibitem{Carli:2010rw}
    T. Carli et al., Eur. Phys. J. C66 503 (2010), [arXiv:0911.2985].
\bibitem{Campbell:2010ff}
     J. M. Campbell and R. K. Ellis, Phys. Rev. D60 113006 (1999), [arXiv:hep-ph/9905386]. \\
     J. M. Campbell and R. K. Ellis, Nucl. Phys. Proc. Suppl. 205–206 (2010) 10, [arXiv:1007.3492].
\bibitem{1405.1067}
  J. R. Andersen et al., (2014), "Les Houches 2013: Physics at TeV Colliders:Standard Model Working Group Report",
  [arXiv:1405.1067], page 80, U.Klein,
"NNLO QCD and NLO EW Drell Yan background predictions for new gauge boson searches". 
  
\bibitem{ref2628of1612.03106}
   S. Catani and M. Grazzini, Phys. Rev. Lett.98 (2007) 222002, [hep-ph/0703012].\\
     S. Catani et al., Phys. Rev. Lett. 103
     (2009) 082001, [arXiv:0903.2120].
   \bibitem{ref2425of161203016}
     R. Gavin et al., Comput. Phys. Commun. 182
(2011) 2388, [arXiv:1011.3540].\\
  R. Gavin et al., Comput. Phys. Commun. 184
(2013) 208, [arXiv:1201.5896].\\
    Y. Li and F. Petriello, Phys. Rev. D86 (2012) 094034, [arXiv:1208.5967].
\bibitem{mrst2004qed}
  A. D. Martin et al., Eur. Phys. J. C39 155-161 (2005), [hep-ph/0411040].
\bibitem{ct10}
   H.-L. Lai et al., Phys. Rev. D82 074024 (2010) , [arXiv:1007.2241].     
 \bibitem{1710.05935}
   R. Ball et al., arXiv:1710.05935
 \bibitem{mcreplicas}
W. T. Giele and S. Keller, Phys.Rev. D58, 094023 (1998), [hep-ph/9803393]\\
W. T. Giele, S. Keller and D. Kosower (2001), [hep-ph/0104052].
 \bibitem{e866}
NuSea Collaboration, R.S. Towell et al., Phys. Rev. D64(2001) 052002, [hep-ex/0103030].  
\bibitem{abm}
S. Alekhin, J. Blümlein, S. Moch, [arXiv:1708.01067].

\end{thebibliography}
\end{document}